\documentclass[notitlepage,twocolumn,pre,tightenlines,superscriptaddress,showpacs,floatfix,nofootinbib]{revtex4-1} 

\usepackage{amsmath,amssymb,amsfonts,latexsym}
\usepackage{ulem}
\usepackage{bbold}
\usepackage{graphicx}
\usepackage{epsfig}
\usepackage{color}
\usepackage[activate=normal]{pdfcprot}  
\usepackage{bm} 
\usepackage{psfrag}
\usepackage[caption = false]{subfig}

\usepackage{hyperref}

\newcommand{\be}{\begin{equation}}
\newcommand{\ee}{\end{equation}}
\newcommand{\ben}{\begin{equation*}}
\newcommand{\een}{\end{equation*}}
\newcommand{\ba}{\begin{eqnarray}}
\newcommand{\ea}{\end{eqnarray}}

\newcommand{\nyuphysics}{Center for Soft Matter Research, Department of Physics, New York University, New York 10003, USA}

\newcommand{\nyusimons}{Simons Center for Computational Physical Chemistry, Department of Chemistry, New York University, New York 10003, USA}
\newcommand{\nyucourant}{Courant Institute of Mathematical Sciences, New York University, New York 10003, USA}

\begin{document}

\title{When you can't count, sample!\\
Computable entropies beyond equilibrium from basin volumes.}

\author{Mathias Casiulis}
\email{mc9287@nyu.edu}
\affiliation{\nyuphysics}
\affiliation{\nyusimons}
\author{Stefano Martiniani}
\email{sm7683@nyu.edu}
\affiliation{\nyuphysics} 
\affiliation{\nyusimons}
\affiliation{\nyucourant}

\date{\today}

\begin{abstract}
In statistical mechanics, measuring the number of available states and their probabilities, and thus the system's entropy, enables the prediction of the macroscopic properties of a physical system at equilibrium. This predictive capacity hinges on the knowledge of the \textit{a priori} probabilities of observing the states of the system, given by the Boltzmann distribution. Unfortunately, the successes of equilibrium statistical mechanics are hard to replicate out of equilibrium, where the \textit{a-priori} probabilities of observing states are in general not known, precluding the na\"{i}ve application of usual tools. In the last decade, exciting developments have occurred that enable the direct numerical estimation of the entropy and density of states of athermal and non-equilibrium systems, thanks to significant methodological advances in the computation of the volume of high-dimensional basins of attraction. Here, we provide a detailed account of these methods, underscoring the challenges that lie in such estimations, recent progress on the matter, and promising directions for future work.
\end{abstract}

\maketitle

\section{Introduction}

From steam engines to LCD displays and polymer materials, the field of thermodynamics has played a pivotal role in the development of modern technology. 
The key to this success has been the efficacy with which equilibrium statistical mechanics predicts the macroscopic properties of physical systems from knowledge of microscopic interactions alone.
This predictive power hinges on the knowledge of the underlying probability distribution for the microscopic states of the system. 
Assuming that the system being considered is in thermal equilibrium, the probability of observing any of its microscopic configurations is given by the Boltzmann distribution, $p_{\mathcal{C}} = Z^{-1} \exp(- \beta E_{\mathcal{C}})$, where $Z=\sum_{\mathcal{C}} \exp(- \beta E_{\mathcal{C}})$ is the partition function, $E_{\mathcal{C}}$ is the energy of the configuration and $\beta$ is an inverse temperature imposed by a thermostat.
Knowledge of the partition function affords us the ability to compute the free energy, $F = - \beta^{-1} \ln Z$, which determines the thermodynamic stability of the state of the system, and allows us to derive all thermodynamic observables from its derivatives.

An alternative, but completely equivalent~\cite{Attard2002}, approach is to compute the entropy of the system,
\begin{align}
    S_G = - \sum_{\mathcal{C}} p_{\mathcal{C}} \log p_{\mathcal{C}}, \label{eq:GibbsEntropy}
\end{align}
and deduce thermodynamics from its maximisation.
This definition of entropy was first introduced by Gibbs~\cite{Gibbs1902} and, when the states of the system occur with equal probability (i.e., $p_\mathcal{C} = 1/\Omega$ with $\Omega$ the volume of the region of accessible states), it reduces to the well-known Boltzmann entropy, $S_B = \log\Omega + const.$~\cite{Kardar2007}. 
The entropy also allows us to predict the direction of spontaneous thermodynamic transformations, as the entropy of an isolated system can only increase with time: put simply, heat flows from hot to cold~\cite{Clausius1854,Carnot1824}.

The state of affairs is not as simple far from equilibrium.
Let us consider an arbitrary system evolving according to some well-defined, but in general non-equilibrium, dynamics.
In such a system, there is no guarantee that objects like temperature or energy can be defined, and there is generally no clear prescription for the probability distribution of microscopic configurations.
Is it still possible to find analogies with equilibrium statistical mechanics?
To address this question, we must introduce two theoretical notions.

First, entropy can be universally defined as the average amount of surprise (or uncertainty) inherent to a random variable (rare observations are more surprising, or informative, than common ones)~\cite{Cover1999}. 
Mathematically, Shannon~\cite{Shannon1948} showed that this amounts exactly to a rewriting of the Gibbs entropy~(\ref{eq:GibbsEntropy}), typically referred to as the Shannon entropy,
\begin{align}
\label{eq:shannon-en}
    S_S = - \sum\limits_{i=1}^{\Omega} p_i \log p_i,
\end{align}
where $p_i$ is the probability of observing state $i$ and the sum is over all accessible states, this time for an arbitrary statistical system with a generic (stationary) probability distribution. 
This information-theoretic interpretation of entropy can be used to define, measure, and interpret entropies in any physical system, whatever its dynamics might be.
This is why there has been recent interest in direct measurements of this quantity using various algorithms, that are not restricted to the one we present here: for instance, the Shannon entropy of physical systems can be estimated using ideas from data compression~\cite{Martiniani2019,Martiniani2020,Avinery2019, cavagna2021vicsek, ro2022model}.

\begin{figure}
\begin{center}
\includegraphics[width=0.46\columnwidth]{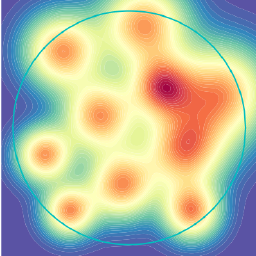} \includegraphics[width=0.46\columnwidth]{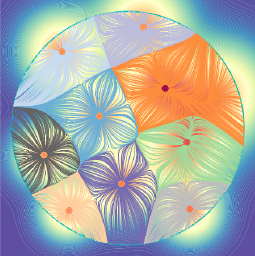}
\end{center}
\caption{\textbf{Illustration of basins of attraction.}
\textbf{Left:} Example of a of $2d$ energy landscape.
High-energy points are shown in blue, low-energy points in red.
\textbf{Right:} In the same landscape, trajectories of steepest descent and ascent $\dot{\bm{x}} = \pm \bm{\nabla} \mathcal{U} (\bm{x})$ are spawned from random points, and confined to an arbitrary region (blue circle).
Trajectories converging to the same minimum belong to the same basin of attraction and are plotted in the same color.} \label{fig:BasinIllu}
\end{figure}

Second, the states of interest of a system can often be described by the stable structures of its dynamics, be they extrema of a high-dimensional function that can be reached by steepest descent (e.g., energy minima of a potential energy landscape), or the fixed points and limit cycles of a generic dynamical system. 
To each of these structures, we can associate a \textit{basin of attraction}, i.e., the set of all initial conditions leading to the same structure via the dynamics, see Fig.~\ref{fig:BasinIllu} for an illustration.

In this perspective we show how we can make use of these facts to arrive to a general protocol for estimating the entropy of systems out of equilibrium, from granular to generic dynamical systems. 
The key observation is that while Shannon's entropy, Eq.~\ref{eq:shannon-en}, provides a universal definition of entropy, we do not know what the a-priori probabilities, $p_i$, of observing a given state of the system are, suggesting that its evaluation may not be possible.
To the contrary, we show that the problem of evaluating these probabilities, $p_i$, or alternatively the problem of enumerating the number of possible states, $\Omega$ (e.g., if one wishes to compute a Boltzmann-like entropy, $S_B=\log \Omega + const.$), can be reduced to a tractable sampling problem.
Indeed, as discussed at length in Sec.~\ref{sec:basinvolume}, one can \textit{compute} the a-priori probability of observing a given stable structure by measuring the volume of its associated basin of attraction~\cite{Martiniani2017} -- and it is therefore possible to estimate the Shannon entropy for an arbitrary system.

In the following, we start by presenting the basin-sampling approach for entropy measurements
in Sec.~\ref{sec:basinvolume}.
Then, in Sec.~\ref{sec:granularentropy}, we highlight its recent successes in the context of granular packings.
Finally, in Sec.~\ref{sec:basinseverywhere}, we discuss exciting extensions of this idea that we believe shall shape the future of computational non-equilibrium physics.

\section{Entropy from basins\label{sec:basinvolume}}

\subsection{Turning intractable counting into sampling}

While the probabilities of observing the states of a nonequilibrium system are in general not known a-priori, it is a generic feature of dynamical systems that for a given initial condition they will end up trapped in small regions of their state space, be they actual basins of an high-dimensional energy landscape, or more general dynamical attractors.
Assuming that there are $\Omega$ such basins, one can readily compute a Boltzmann-like entropy, $S_B = \log\Omega + const.$
In a thermodynamic system, it simply amounts to the entropy at zero temperature.
If we restrict ourselves to a set of typical inherent structures observed at finite temperature (viz., the configurations that the system relaxes to by steepest descent from finite temperature samples), this quantity becomes the system's configurational entropy (distinct from the vibrational contributions to the total entropy)~\cite{Sciortino2005}.
By analogy, in any dynamical system, one can always define such an entropy, and compute it by counting basins of attraction.

Counting the number of basins of attractions is tantamount to enumerating all microstates in the microcanonical ensemble of statistical mechanics: while it is in principle a valid strategy, it is in practice numerically intractable.
Until recently, this issue was believed to be impossible to overcome ~\cite{Wiley2006, Cornelius2013, Zhang2021}.
The problem has since been solved by the introduction of Monte Carlo methods capable of integrating the volume of individual basins of attraction using ideas originally introduced for the calculation of the free energy of solids in statistical mechanics~\cite{Frenkel1984}.
Thanks to this new class of methods, pioneered by Frenkel, Xu~\cite{Xu2011}, Asenjo ~\cite{Asenjo2014}, and Martiniani~\cite{Martiniani2017a, Martiniani2016, Martiniani2016a, Frenkel2017}, alongside collaborators, it is now possible to approach these problems and measure quantities that could never have been computed with previous techniques.

The gist of the method is as follows: since the accessible volume, $\mathcal{V}$, of configuration space is tiled by $\Omega$ basins of attraction, by definition of the arithmetic mean we have that
\begin{align}
    \mathcal{V} = \sum\limits_{i=1}^{\Omega} v_i = \Omega \langle v \rangle,
    \label{eq:mbv}
\end{align}
where $v_i$ is the volume of basin $i$ in configuration space, and $\langle \cdot \rangle$ is the mean taken over all basins.
As a result, one can write the number of basins as $\Omega =  \mathcal{V} / \langle v \rangle$.
This seemingly innocent equation is in fact crucial, as it turns the intractable enumeration problem of counting $\Omega$ into a sampling problem, namely the computation of the average basin volume $\langle v \rangle$, which can be performed over a finite set of basins.

There are two main challenges in computing this quantity in the context of many-body systems: on the one hand, high-dimensional geometry makes volume estimations difficult and, on the other hand, mean volume estimations by direct sampling are typically biased. 
We start by considering the challenges of estimating volumes in high dimensions, and how we can overcome them by means of suitably-modified free energy calculations.

\subsection{A starter in high-dimensional geometry: there is plenty of room in the corners}
\begin{figure*}
\begin{center}
\includegraphics[width=0.48\columnwidth]{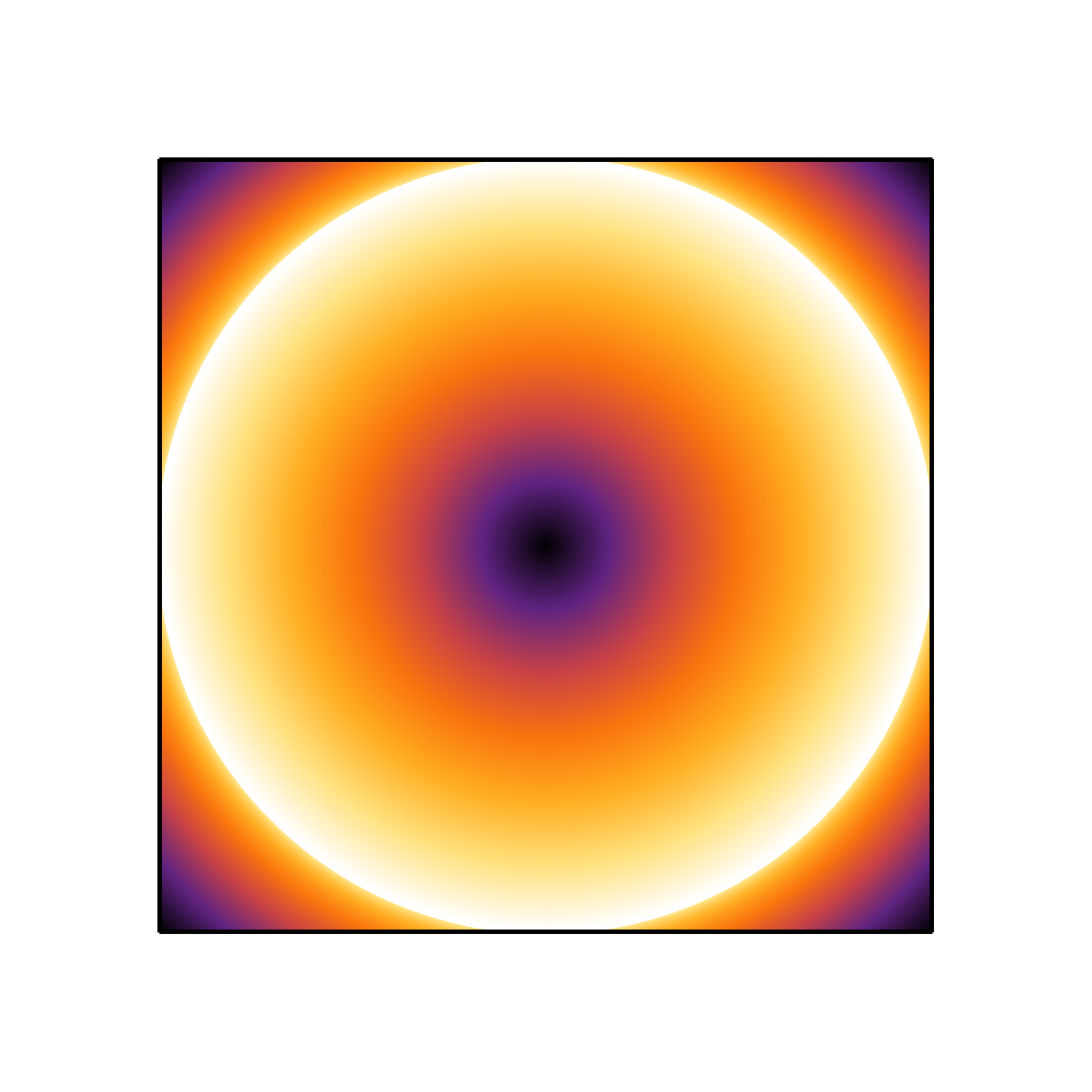}
\includegraphics[width=0.48\columnwidth]{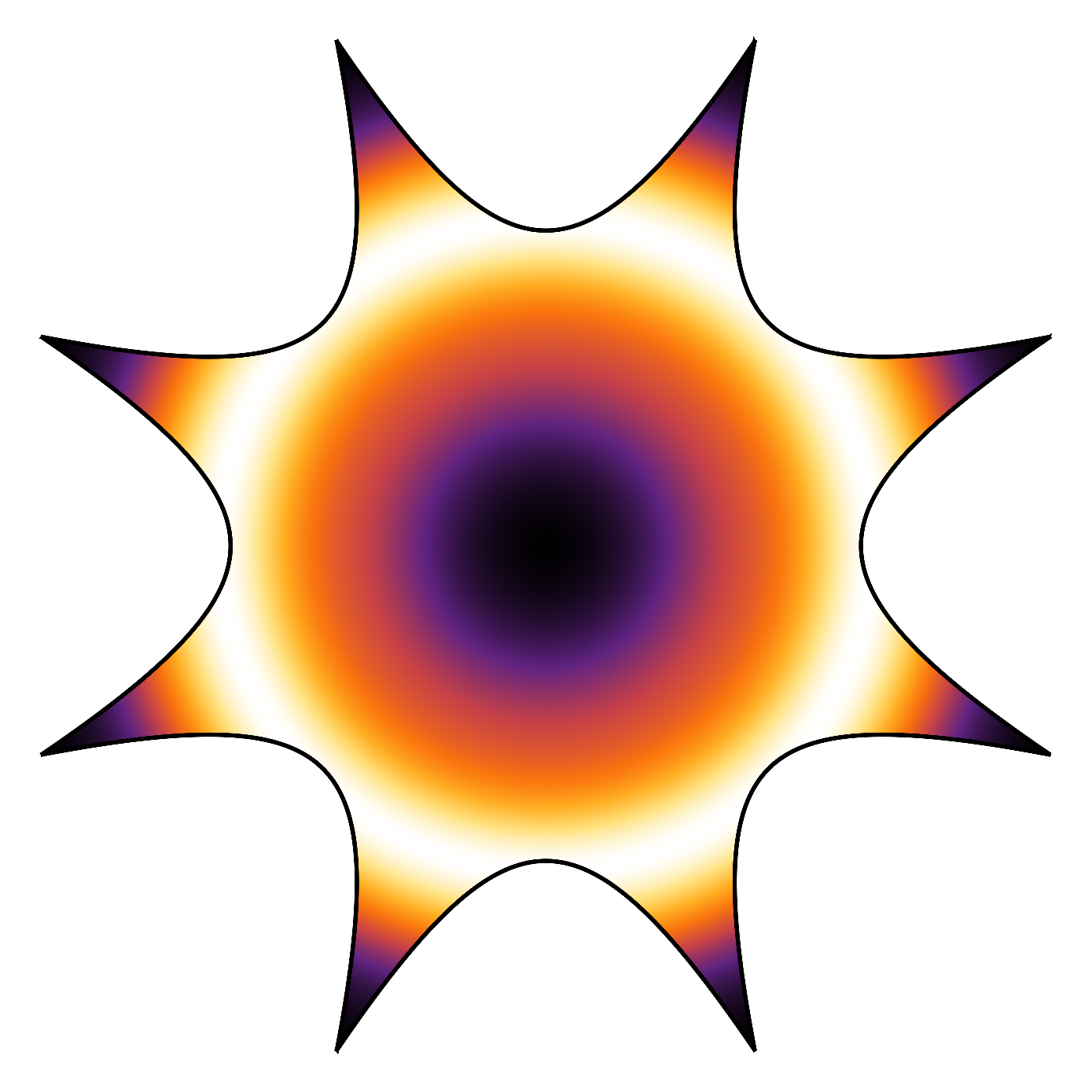}
\includegraphics[width=0.48\columnwidth]{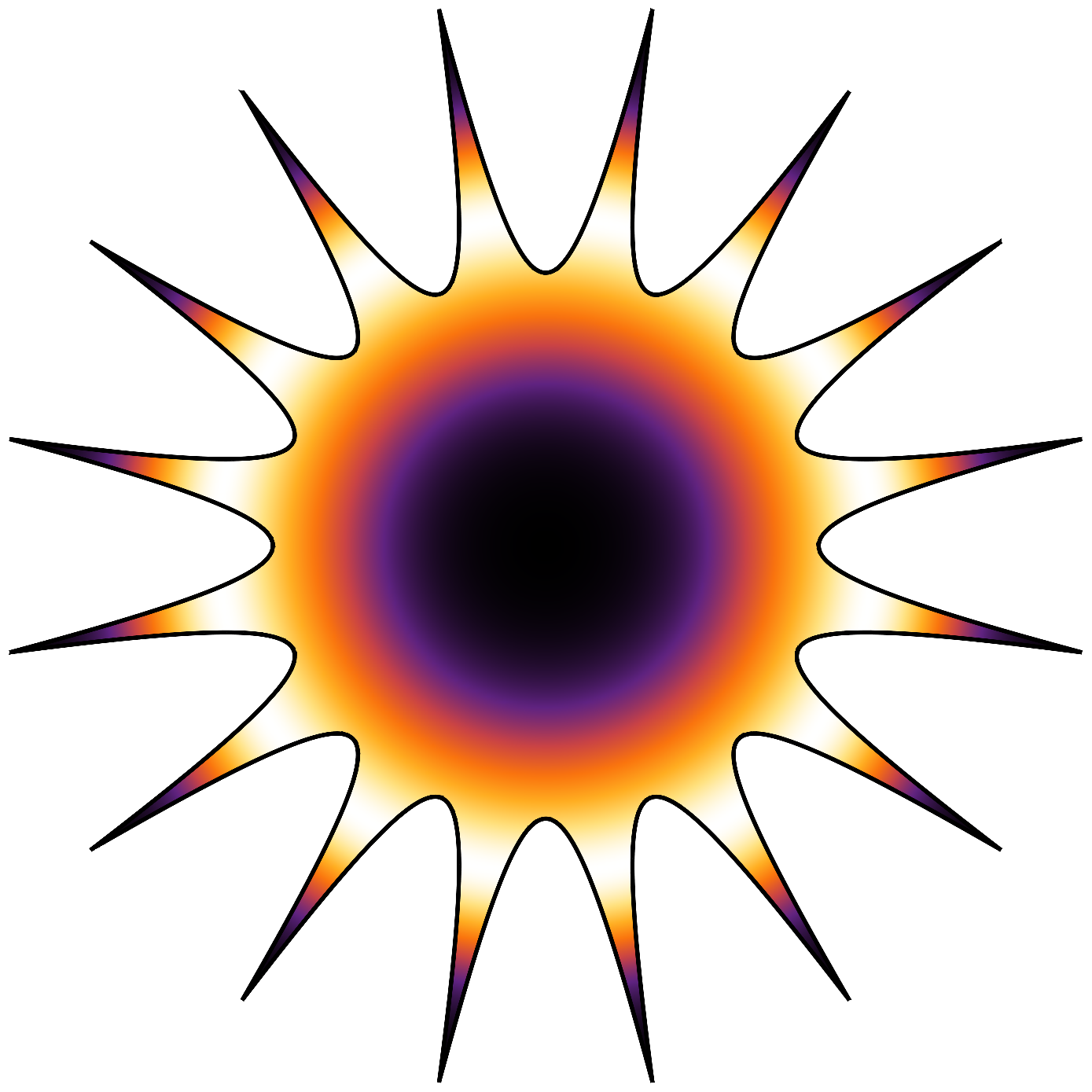}
\includegraphics[width=0.48\columnwidth]{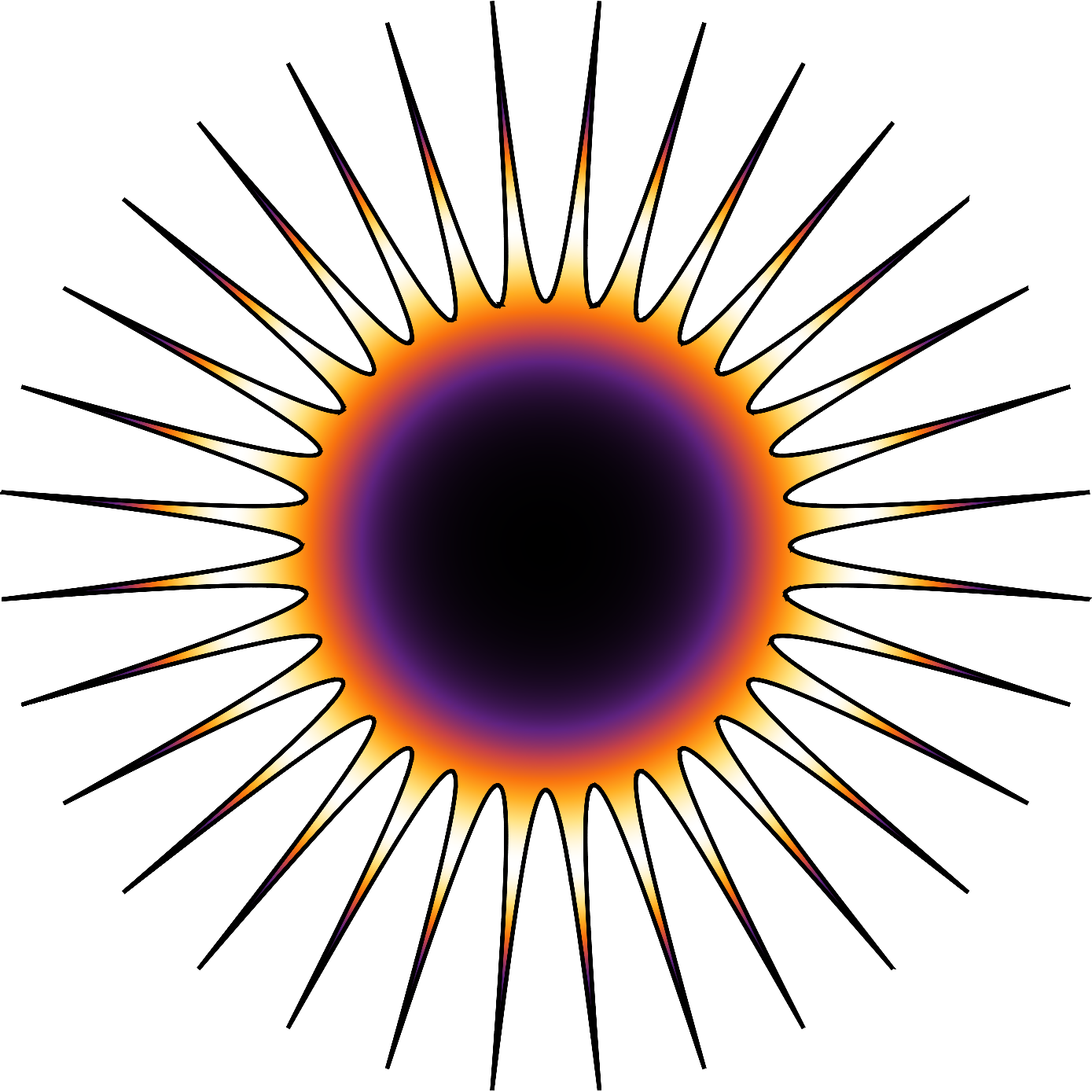} \\
\hspace{0.003\columnwidth}
\includegraphics[height=0.42\columnwidth]{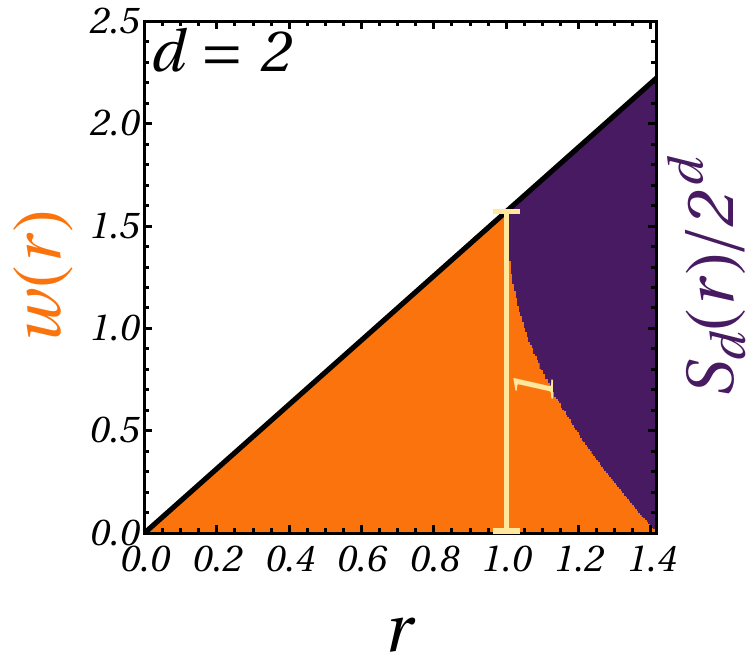}
\hspace{0.000\columnwidth}
\includegraphics[height=0.42\columnwidth]{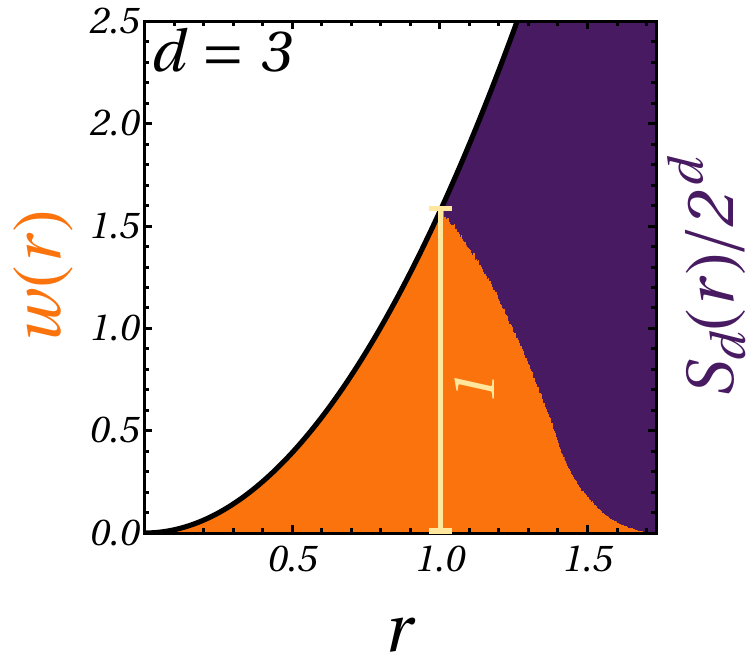}
\hspace{0.000\columnwidth}
\includegraphics[height=0.42\columnwidth]{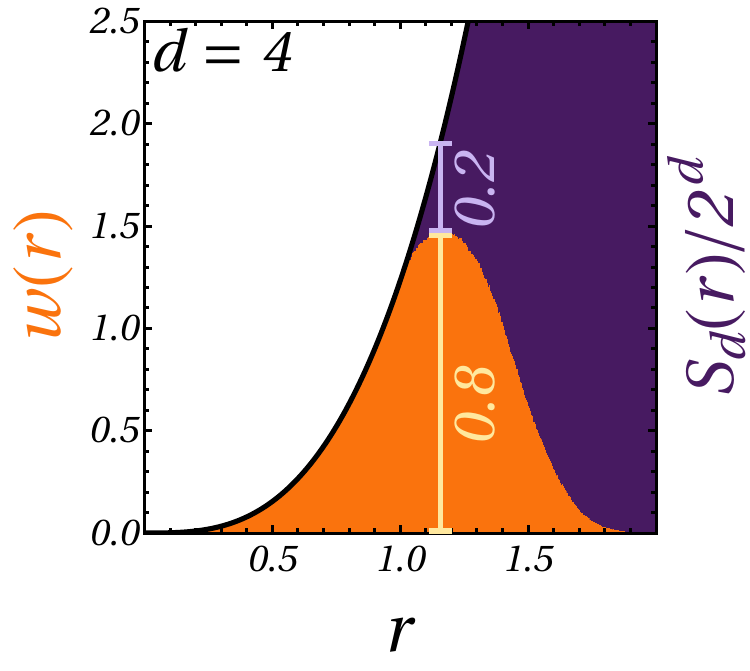}
\hspace{0.000\columnwidth}
\includegraphics[height=0.42\columnwidth]{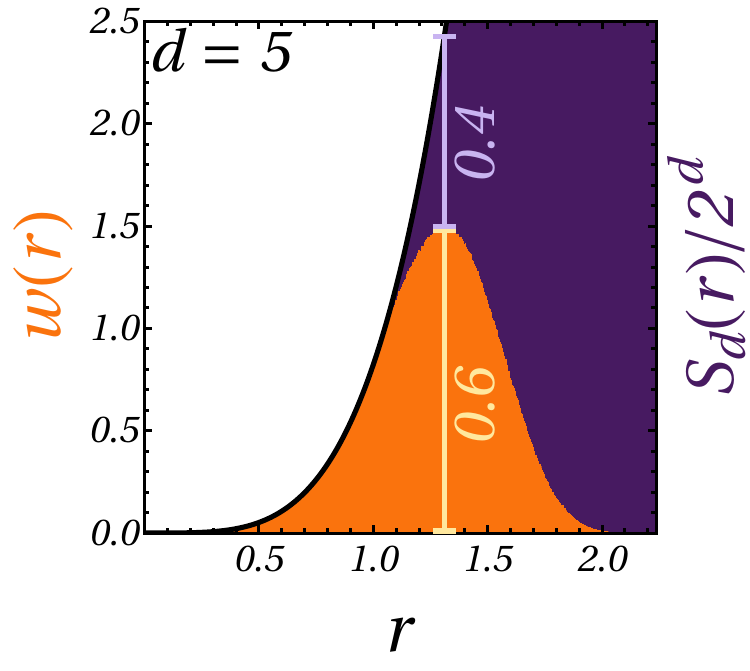}
\hspace{0.004\columnwidth}
\end{center}
\caption{\textbf{High-dimensional volumes.}
\textbf{Top:} $2d$ sketches of hypercubes in $d = 2, 3, 4, $ and $5$.
The outer black line links vertices located at a distance $\sqrt{d}$ from the center of the cube to midpoints located on the unit hypersphere with hyperbolic lines.
The inside of each shape is colored with a density plot of the relative contribution $w(r)$ of each radial shell inside the cube to the full volume, as obtained from Monte Carlo integration using $10^6$ points uniformly drawn inside each cube.
\textbf{Bottom:} Histograms of relative weights of radial shells $w(r)$ for the hypercube (orange) and for the hyperball (purple).
The lines highlight how large the intersection of the cube and a hyperspherical shell is at the distance that contains most of the cube's volume.
As reminded in the insets, this intersection here represents $1, 1, 0.78,$ and $0.62$ times the surface of the full hyperspherical shell, respectively.
} \label{fig:HighdimIllu}
\end{figure*}
The computation of high-dimensional basin volumes is a difficult task that cannot be accomplished by simple quadrature~\cite{Wiley2006}.
To illustrate this difficulty, we take the example of a simple shape with a volume that is known in any dimension: the hypercube.
In the top line of Fig.~\ref{fig:HighdimIllu}, we present $2d$ sketches of the aspect of hypercubes, $\bm{r} \in \left[-a;a\right]^d$, in dimensions $d = 2,3,4,$ and $5$.
In each case, following recommendations on drawings of high-dimensional convex volumes~\cite{Artstein-Avidan2015}, we represent the $d-$dimensional hypercube with half-sidelength $a=1$ by linking each of the $2^d$ summits, that sit at a distance $\sqrt{d}a$ from the center, to $2$ of its neighbors using hyperbolas tangent to the largest hypersphere contained by the hypercube -- namely the unit hypersphere.
The reason for choosing hyperbolas is that if one splits the cube into two radial regions of equal volumes, $\mathcal{R}_{\textrm{in}}= \left\{r:  r<R_{1/2} \right\}$ and $\mathcal{R}_{\textrm{out}} = \left\{ r: r>R_{1/2}\right\}$, where $r$ is the distance from the center and $R_{1/2}$ a threshold distance, the $(d-1)$-dimensional cross-section of each diagonal of the hypercube decays exponentially with $r$ in $\mathcal{R}_{\textrm{out}}$~\cite{Artstein-Avidan2015}.
In other words, in large $d$, one corner of a hypercube has a vanishing volume over surface ratio, a property shared with hyperbolic objects, like Gabriel's horn (also called Torricelli's trumpet) in $3d$~\cite{MathworldGabrielsHorn}.

This is an illustration of the fact that, in high dimensions, even simple compact objects present tendril-, or tentacle-like regions~\cite{Ashwin2012, Martiniani2016a, Martiniani2017, Martiniani2016, Zhang2021} that extend very far but get very thin.
Yet, somewhat counter-intuitively, most of the volume of high-dimensional objects is contained in these extended objects!
This is shown in Fig.~\ref{fig:HighdimIllu} by the density plot inside each shape, that represents the actual weight of each hyperspherical shell within each hypercube.
While in $d=2$, the shell that contributes the most to the volume is the unit circle, as the dimensionality increases, it shifts to larger and larger radial distances, at $R_{max} \sim \sqrt{d/3}$ in the limit of large $d$.\footnote{
This can be shown by a simple statistical argument: suppose that we uniformly draw a random vector within a $d-$dimensional hypercube, with coordinates $(x_1,\ldots,x_d)$. The squared distance of this point from the center, $r^2 = \sum x_i^2$ is a sum of independent, identically drawn uniform random variables so that, per the central limit theorem, its distribution tends to a Gaussian with mean value $\mu = d \langle x^2 \rangle = d/3$, which asymptotically yields $\mathbb{E}[\sqrt{r^2}]\sim \sqrt{d/3}$.}

To complete this picture, in the second row of Fig.~\ref{fig:HighdimIllu}, we plot in orange the distribution of mass of hypercubes with $a=1$ along the radial direction, $w(r)$, and in purple the corresponding line for the hyperball, which is simply the surface area $S_d$ of the $(d-1)$-sphere with radius $\sqrt{d}a$, normalised like $w(r)$, \textit{i.e.} by the volume of the cube, $(2a)^d$.
As the dimensionality of space increases, the volume of the hypercube concentrates deeper and deeper into disconnected tendril-like objects, at distances such that the intersection between the hypercube and the hypersphere becomes very small: in Fig.~\ref{fig:HighdimIllu}, as highlighted by lines and inset texts, the ratio between the value of the orange curve at its maximum and the value of the purple curve at the same distance decreases with $d$.
Furthermore, the contribution of the unit ball to the total volume of the hypercube vanishes as the dimension increases.
These last two properties make practical volume estimations extremely complicated: a na\"{i}ve Monte Carlo integration using uniformly sampled points in a hyperball is bound to fail since the vast majority of points will fall outside of the shape of interest, while an integration within a ball close to the center will yield only a tiny fraction of the overall volume, an example of the so-called curse of dimensionality~\cite{Ashwin2012, Martiniani2016a, Zhang2021}.

\begin{figure}
\begin{center}
\includegraphics[width=0.96\columnwidth]{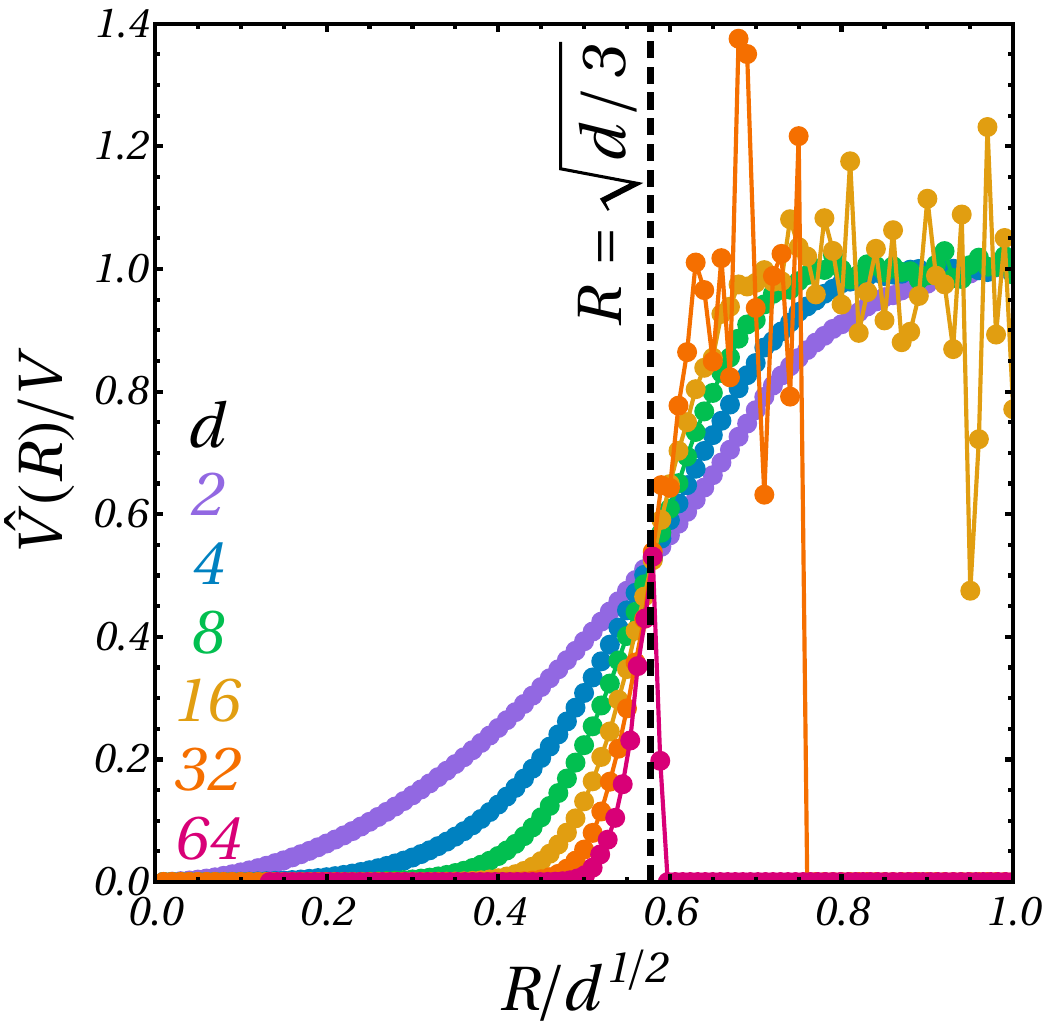}
\end{center}
\caption{\textbf{Na\"{i}ve random sampling shortcomings.}
Direct Monte Carlo evaluation $\hat{V}(R)$ of the volume of a hypercube with actual volume $V$, using $N_s = 10^5$ points uniformly drawn from a hyperball with radius $R$, for dimensions going from $2$ to $64$.
Here the half-sidelength of the hypercube is $a=1$.
The dashed line indicates the asymptotic mode of the mass distribution of the cube, $\sqrt{d/3}$.
} \label{fig:HighdimSamplingFailures}
\end{figure}

To more concretely illustrate how na\"{i}ve random sampling methods fail at estimating volumes in high dimensions, let us consider the volume estimation problem for a simple hypercube, using Monte Carlo integration.
We sample $N_s$ points uniformly drawn in a hyperball, $\mathcal{B}(R)$, with radius $R$ centered at the same point as the hypercube, measure the fraction $f_{MC}$ of points that fall within the hypercube, and compute the corresponding volume, $\hat{V}(R) = f_{MC} V_{\mathcal{B}}(R)$.
In general, one does not necessarily know the largest linear size of the object of interest, so that $R$ should in principle be varied.
We use this strategy to estimate the volume of hypercubes with $a = 1$ by varying the ball radius, $R$, between $0$ and the length of the longest diagonal of the cube, $\sqrt{d}$.
In Fig.~\ref{fig:HighdimSamplingFailures}, we plot the measured volume divided by the true volume of the hypercube, $V = 2^d$, against $R/\sqrt{d}$, for $d$ between $2$ (mauve) and $64$ (red), for $N_s = 10^5$.
At low dimensions of space, this method converges smoothly to $V_{MC} = V$ as $R \to \sqrt{d}$.
As $d$ increases, the volume concentrates more and more around the mode of the distribution of mass of the cube, $\sqrt{d/3}$, and as a result the curves become step-like.
However, as $d$ increases, the measurement also becomes less reliable upon approaching $R \to \sqrt{d}$: it first displays increasingly large fluctuations ($d\leq16$) then violently falls to $0$ ($d\geq 32$).
The reason is that most points in the ball are actually sampled at radii such that the hypercube is already made of narrow spikes, as illustrated in Fig.~\ref{fig:HighdimIllu}.
Indeed, the ratio of the volume of the hypercube to that of the hyperball with radius $\sqrt{d}$ decays exponentially with dimension, $V/V_{\mathcal{B}}(\sqrt{d}) \sim \sqrt{d}\exp(-d)$.

If we did not know that we were measuring a simple cube, and ball-picked from spheres of increasing radius to produce the same kind of curve as in Fig.~\ref{fig:HighdimSamplingFailures}, in high dimension our estimate would be far off as the mode of the mass distribution of the cube moves further and further into the corners.
In the simple example here, in $d=64$, the best estimate with our choice of $N_s$ would be roughly $50\%$ off!
This shows that simple Monte Carlo integration, while efficient in small dimensions, fails at capturing volumes in high dimensions.
Note that the case illustrated here is actually fairly ideal: we know the location of the center of the hypercube, which makes the measurement easier, and the hypercube is a rather regular object.
In real-world situations, a Monte Carlo integration of high-dimensional volumes is likely to be far worse.

\subsection{Free energy methods for volume computations}
\begin{figure*}
\begin{center}
\includegraphics[width=0.48\columnwidth]{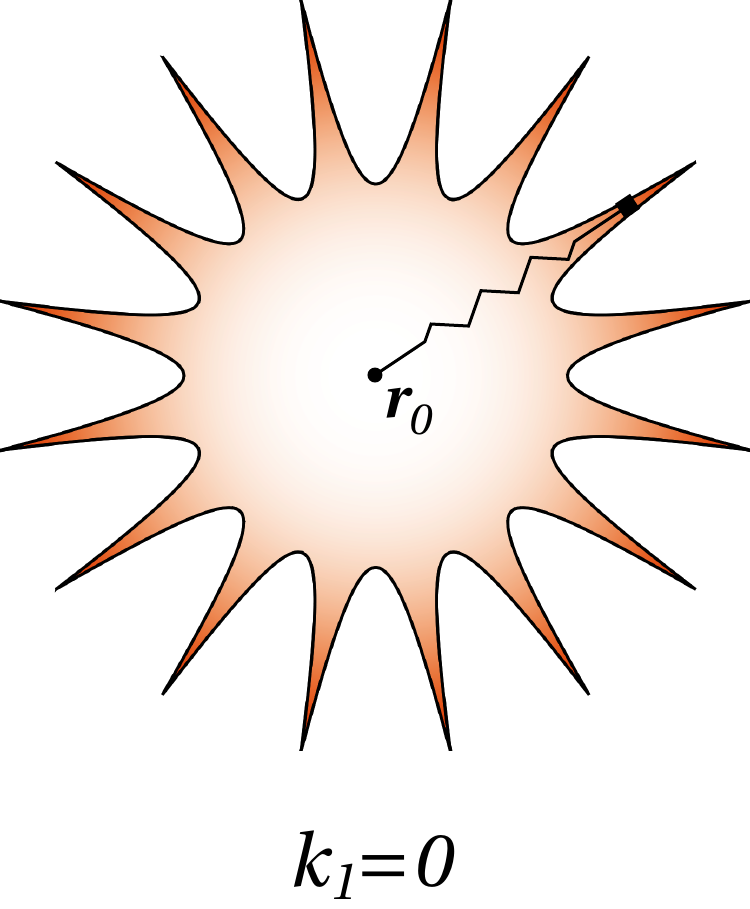}
\includegraphics[width=0.48\columnwidth]{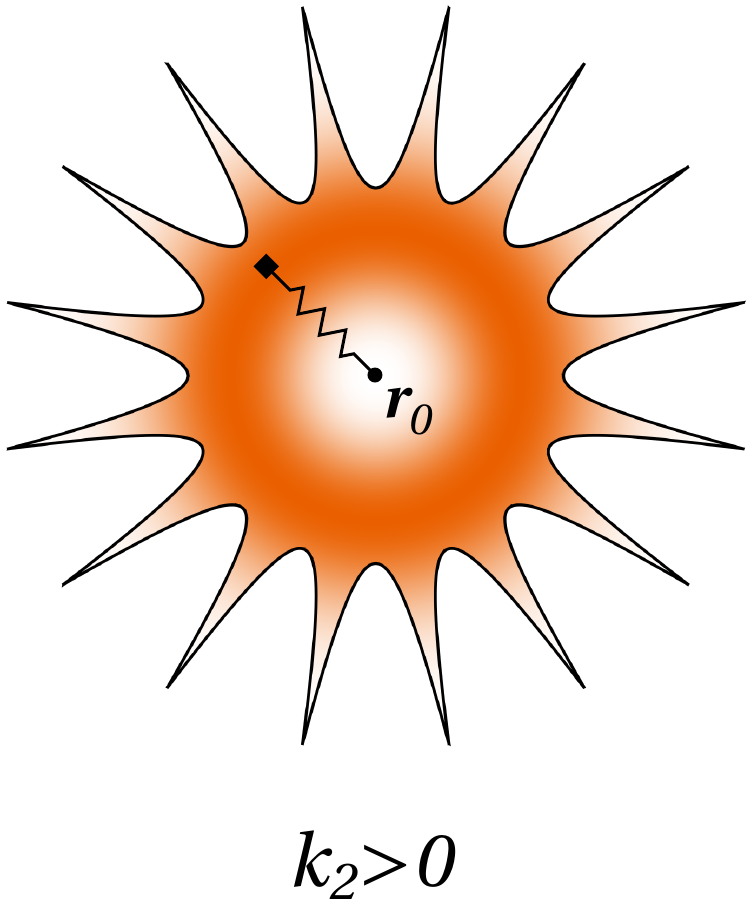}
\includegraphics[width=0.48\columnwidth]{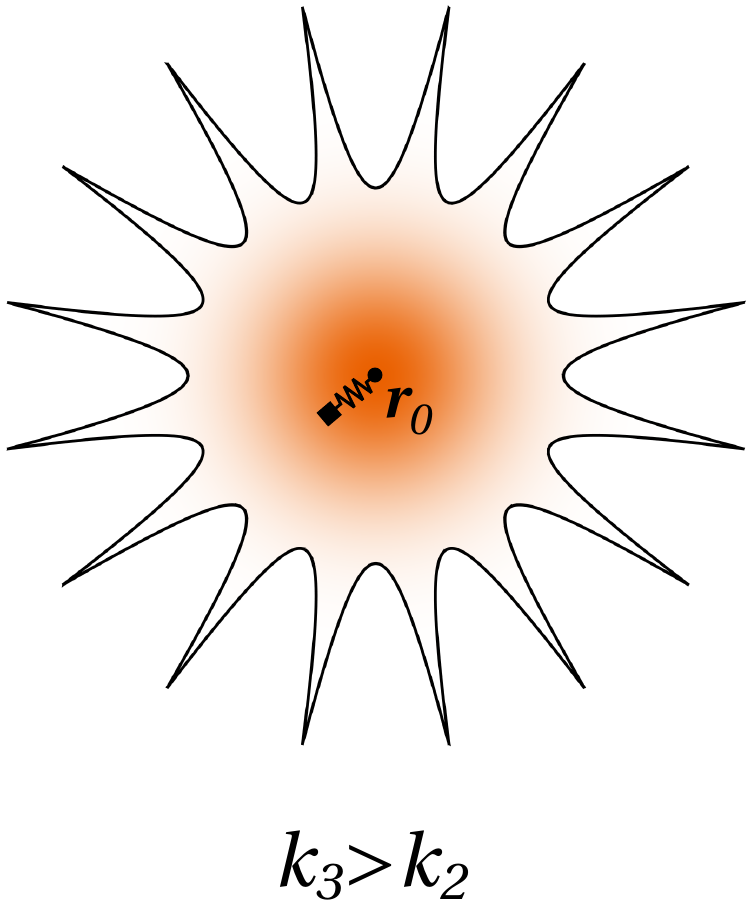}
\includegraphics[width=0.48\columnwidth]{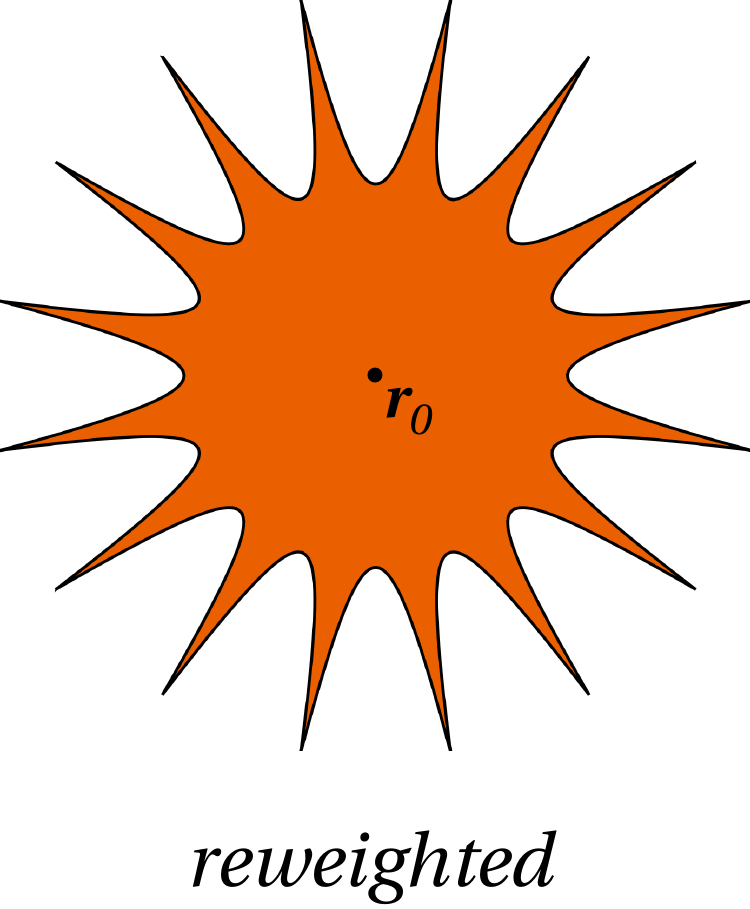}
\end{center}
\caption{\textbf{Probing basin volumes with random walks.}
Different random walkers are linked by springs with different rigidities to the configuration at the minimum.
The first three panels represent the probability density of the trajectories of such walkers, with rigidities growing from left to right.
At very small (possibly negative) rigidities, the walker explores the outer rim of the basins, following tentacle-like structures, while at large rigidities the random walk typically explores a hypersphere around the minimum.
Putting together and properly reweighing the information sampled by these walkers, one can faithfully recover the full volume as if it had been uniformly sampled (last panel).
} \label{fig:MCIllu}
\end{figure*}
It has been shown that volume estimations and enumeration problems can be treated in a way that amounts to a free energy calculation in the spirit of the Frenkel-Ladd method~\cite{Frenkel1984}.
The idea is that the volume $v$ of a domain $\Gamma$ can be rewritten as
\begin{align}
    v = \int_{\mathbb{R}^d} \mathrm{d}\bm{r} \mathcal{O}_{\Gamma}(\bm{r}) = \int_{\mathbb{R}^d} \mathrm{d}\bm{r} e^{-\beta \mathcal{U}_{\Gamma}(\bm{r})},
\end{align}
where $\mathcal{O}_\Gamma $ is the characteristic function of the domain to integrate over, or ``oracle'', which can be exponentiated to yield the potential $\mathcal{U}_{\Gamma}$ which is $0$ inside the domain and $\infty$ outside it.
Written in this form, the volume can be interpreted as the partition function of a free Brownian walker exploring the domain $\Gamma$, with a hard wall at the boundary, at an inverse temperature $\beta$.

The most popular and earliest class of methods for the computation of partition functions is based on thermodynamic integration (TI)~\cite{Kirkwood1935,Gelman1998,Frenkel2001}, which consists of parameterizing the system's Hamiltonian in such a way that we can ``morph'' an unknown partition function into one that we know how to compute analytically.
In practice, this amounts to replacing a high-dimensional integral over phase space volume with a low-dimensional integral over one or more Hamiltonian parameters (each point in the integrand is obtained from an equilibrium simulation with a given choice of Hamiltonian parameters), or to estimating ratios of partition functions from equilibrium samples obtained from simulations with different Hamiltonian parameters~\cite{Bunker2000, Fukunishi2002}, as we show presently.

In the context of volume estimations, the potential $\mathcal{U}_\Gamma$ that encodes the oracle is athermal, so that we can take $\beta=1$ without loss of generality.
Then, in the spirit of umbrella sampling~\cite{Torrie1977,Frenkel2001}, we can introduce simple biasing potentials that depend on a control parameter that allows us to go continuously from an integral of unknown volume to one of known volume.
For instance, one can tether random walkers to a reference point, $\mathbf{r}_0$, inside the basin of attraction (e.g., the minimum energy configuration) using harmonic springs of varying stiffness $k$,  see Fig.~\ref{fig:MCIllu}.
For a random walk constrained to remain within the domain of interest $\Gamma$, one can compute a volume $v_k$ weighted by the Boltzmann factor (viz., the corresponding partition function) as
\begin{align}
    v_k &= \int_\Gamma \mathrm{d}\bm{r}e^{-\frac{1}{2}k |\mathbf{r}-\mathbf{r}_0|^2} \label{eq:Gaussian}
\end{align}
and define the \textit{dimensionless basin free energy} as the negative log-volume $f_k = - \log v_k$. 
When $k = 0$, the walker is completely free to explore the basin volume, while for $k\to \infty$ the walk is reduced to a small region surrounding $\mathbf{r}_0$ that fits entirely within the basin.
Computing the basin volume amounts to measuring the dimensionless free energy difference between the walkers with $k = 0$ and $k \to \infty$ so that
\begin{align}
    f_{k=0} = f_{k\to\infty} + (\hat{f}_{k=0} - \hat{f}_{k\to\infty}) \label{eq:FEdiff}
\end{align}
where we added the analytical reference free energy $f_{k\to\infty}$ to the numerical estimate (denoted by a “hat”) of the free energy difference between $k = 0$ and $k \to \infty$.
This is necessary because free energies can only be computed numerically up to an additive constant equal for all $k$’s.
The analytical reference can be computed by a Gaussian integral at the largest stiffness value, $k_{max}$, provided that $k_{max}$ is large enough that the corresponding random walk is essentially unaffected by the boundary of the domain.

Intuitively, this importance sampling method should outperform brute-force Monte Carlo sampling because the steps of the walks are chosen to remain close to one another, so that even within the high-dimensional tentacles, the rejection rate of the steps will be much smaller than what we would get by ball-picking within a sphere that contains the shape of interest.
In other words, this method samples points compactly within the volume being estimated, rather than throwing darts at random in a huge volume around it. 

Nevertheless, running independent biased random walks remains a poor strategy in high dimensions.
Take for instance the example of the hypercube: Due to the $r^{d-1}$ scaling for the surface of a hypersphere, most points of the walk will be concentrated far away from the center of the cube, and instead will live close to the heaviest shell of the mass distribution at $\sqrt{d/3}$.
In high dimensions, this maximum lies far into the corners, so that a single random walk would typically spend very long times in one of the $2^d$ corners.
As a result, attempting a direct estimation of $v$ with independent free random walks would typically require either exponentially many ($2^d$) realizations, or exponentially long equilibration times so that the walk can escape a given corner, reach the (typically tiny) convex core, and explore another corner, $2^d$ times.

To overcome this problem, one can take inspiration in the large body of work on free energy estimations at low temperature in rugged energy landscapes (see for instance~\cite{Frenkel2001,LandauBinder}).
A typical strategy is to use parallel tempering ~\cite{Martiniani2016, Bunker2000, Fukunishi2002}, which amounts to running a collection of simulations with different control parameters (typically different temperatures), each called a ``replica'', and allowing for configuration exchanges between high and low temperature replicas so that the low temperature replicas do not become trapped in a local region of the energy landscape, all while respecting the detailed balance condition.
In the context of basin volume calculations, we exchange coordinates between random walks with different stiffness, $k$, so that walkers at low stiffness can escape the tentacles by swapping coordinates with walkers at high values of $k$ that are constrained to “live” closer to the hyperspherical core, as illustrated in Fig.~\ref{fig:MCIllu}.

The dimensionless free energy difference in Eq.~\ref{eq:FEdiff} can be computed by a simple thermodynamic integration over $k$~\cite{Frenkel2001} or using more sophisticated estimators like the Multi-Bennet Acceptance Ratio Method (MBAR)~\cite{Shirts2008}, which should yield more statistically accurate results.
The MBAR estimator hinges on the idea that one can always relate the free energy at one value of $k$ with the free energy at every other value of $k$, via
\begin{align}
    \hat{f}_k = - \ln \left[ \sum\limits_{m = 1}^{K} \sum\limits_{a = 1}^{N_m} \frac{\exp[- \beta \mathcal{U}_k(\bm{r}_a)]}{\sum\limits_{k'} N_{k'} \exp[\hat{f}_{k'} - \beta \mathcal{U}_{k'}(\bm{r}_a)]}\right],
\end{align}
where $\mathcal{U}_k(\mathbf{r})=-k|\mathbf{r}-\mathbf{r}_0|^2/2$ is the biasing potential with spring stiffness $k$ (but can in principle assume any shape), $K$ is the total number of biased random walks, and $N_m$ is the number of uncorrelated equilibrium samples obtained from the $m$-th random walk.
This system of implicit equations can be solved numerically, typically using a self-consistent Newton-Raphson scheme~\cite{Shirts2008}.
Ideally, for the quality of the MBAR solution to be as good as possible, one should measure a similar amount of equilibrium samples in every region of the volume being measured.
This problem boils down to the choice of $\mathcal{U}_k$ and we discuss it briefly in the next subsection.

Finally, MBAR yields the optimal reweighing of the set of histograms $h_k(r)$, where $r\equiv|\mathbf{r}-\mathbf{r}_0|$, allowing for the reconstruction of the full density of states, which, in the context of volume measurements, amounts to the mass distribution, $h(r)$, within the object of interest, that verifies
\begin{align}
    \ln h(r) = \sum_{k} w_k(r) \left[ \ln h_k(r) + \beta \mathcal{U}_k(r) - (\hat{f}_{k} - \hat{f}_{0})\right], \label{eq:MBARHistos}
\end{align}
with $w_k(r) = h_k(r) / \sum_{k'} h_k'(r)$ a set of normalised weights.

\subsection{Measuring hypercubes}

To illustrate how effective this technique is compared to brute-force Monte Carlo, we now apply it to the measurement of high-dimensional hypercubes, $\bm{r} \in \left[-a;a\right]^d$.
For each dimensionality, $d$, we run a collection of $K_d$ random walks -- that we refer to as replicas -- associated to spring constants $(k_1 = 0, k_2, \ldots, k_{K_d})$, for $N_s = 5 \times 10^5$ steps, and attempting a parallel tempering coordinate exchange between replicas with neighboring values of $k$ every $10$ steps.
The number of replicas at a given dimensionality, $K_d$, is set so that the histogram of sampled distances from the center, $h_{k_i}(r)$, overlaps significantly with the histogram of replicas with neighboring values of $k$, namely $h(r)_{k_{i \pm 1}}$.

The scaling of the number of required replicas with dimensionality, $d$, can be estimated from the mean and variance of the radial distribution for a $d$-dimensional Gaussian, $h_k(r) \sim r^{d-1}\exp{(-kr^2/2)}$.
We start by noting that for parallel tempering to be efficient, and for the MBAR estimation to be reliable, these distributions must display significant overlap.
Choosing the replicas so that their modes are separated by the standard deviation of the narrowest walk, one can show that the required number of replicas, $K_d$, scales linearly with $d$ (see App.~\ref{app:replicaspacing}). 
Recall, for context, that na\"{i}ve Monte Carlo sampling typically requires a number of samples that grows \textit{exponentially} with $d$.

\begin{figure}[b]
    \centering
    \includegraphics[width=0.96\columnwidth]{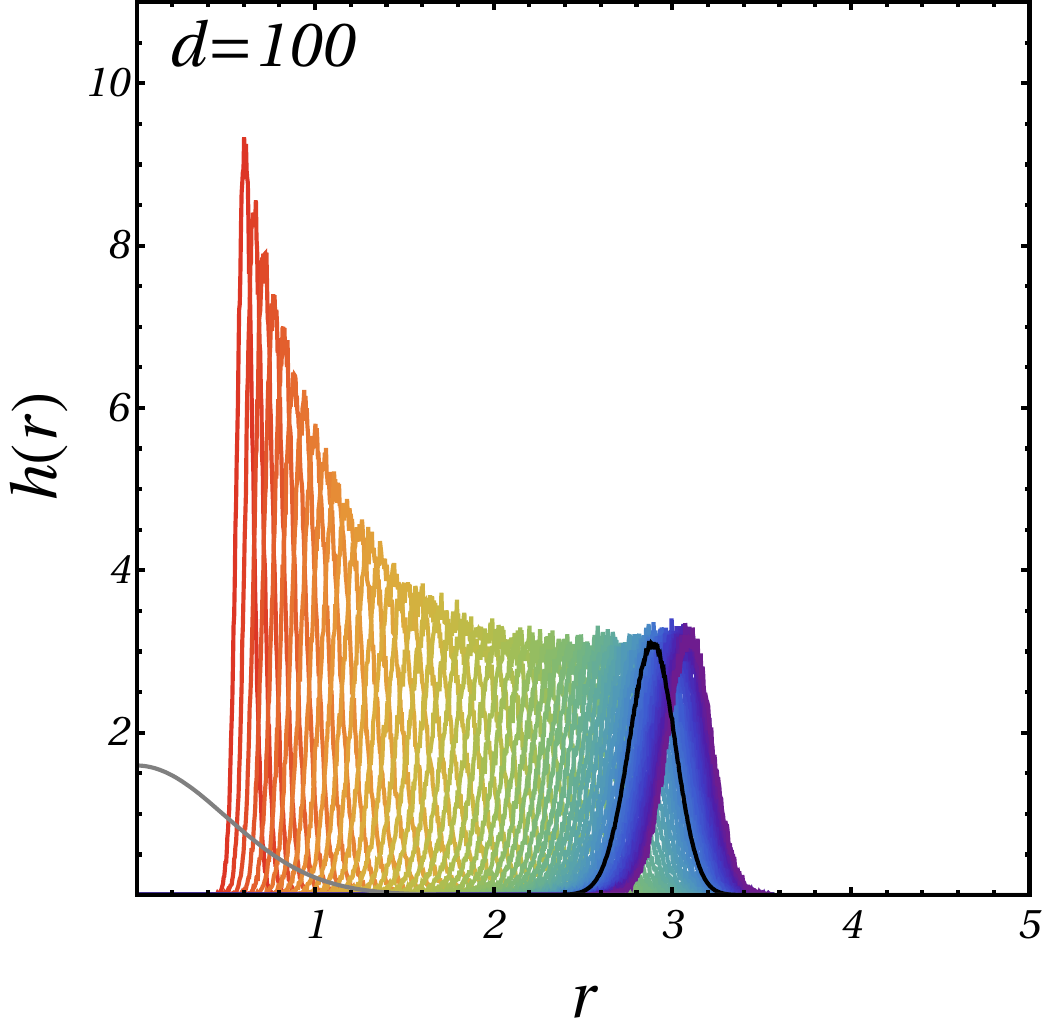}
    \caption{\textbf{Biased walks in a hypercube.} 
    Each colored line is a histogram $h_k(r)$ of distances to the center for a random walk constrained to remain in a $100d$ hypercube with unit sidelength, $a = 0.5$, and subjected to a biasing potential $\mathcal{U}_k(r) = k r^2/2$.
    Color encodes the value of the rigidity $k$, from large (red) to small (purple).
    Here, the $K = 64$ spring constants are chosen so that the modes of the distribution are equally spaced in $r$, from $r = a = 0.5$ (red) to $r \approx 3.1 > \sqrt{d/3}a$ (purple).
    A $1d$ gaussian is sampled to obtain points near the center (gray line) to improve the MBAR solution.
    The reconstructed density of states of the cube is shown in black.
    Note that the $16$ histograms to the right of the black line correspond to negative values of $k$.
    }
    \label{fig:CubeHistos}
\end{figure}
We show the output histograms, $h_k(r)$, for a practical implementation of these random walks in a $d=100$-dimensional hypercube in Fig.~\ref{fig:CubeHistos}.
In this example, we used $K = 64$ replicas with $k$'s chosen so that the modes of the distributions are linearly spaced in $r$.
Of these $64$ $k$ values, $48$ were positive, leading to $k=0$, and $16$ were negative (the most negative being $k=-4.5$), so that the biasing potential pushes these walkers further into the corners of the cube.
As expected, high values of $k$ (red histograms) yield narrower distributions closer to the origin, while lower ones (towards purple) are broader and distant from the origin. 

We chose $k_{max}$ (the left-most curve in Fig.~\ref{fig:CubeHistos}) 
so that its mode would coincide with that of the largest ball inscribed in the cube.
Note how this leaves a significant gap in the histograms in the regions nearest to the origin due to the power-law $r^{d-1}$ in $h_k$ (yet another manifestation of the curse of dimensionality).
In order to improve the MBAR free energy estimation, and the density of states reconstruction, we chose to sample independent points from $h_{in}(\bm{r}) \propto r^{1-d} e^{-k_{in} r^2/2}$ , also constrained to the inside of the cube, to accumulate samples near the origin (gray line in Fig.~\ref{fig:CubeHistos}).
The corresponding biasing potential is $U_{in} = -(d-1)\ln r + k_{in} r^2/2$, where we choose $k_{in} = 4$ so that the distribution has width $\sigma \sim 1/\sqrt{k_{in}} = 0.5$.

Finally, to get absolute free energy values, we use the negative log-volume of the largest ball inscribed in the hypercube as the known reference in Eq.~\ref{eq:FEdiff}, instead of the free energy for a tight harmonic trap, so that the free energy becomes
\begin{equation}
    f_{k=0} = f_{\mathcal{B}(a)} + (\hat{f}_{k=0} - \hat{f}_{\mathcal{B}(a)})
\end{equation}
where $\hat{f}_{\mathcal{B}(a)}$ is the estimated free energy for a set of points sampled directly from the largest inscribed ball, $\mathcal{B}(a)$, whose volume is known analytically.
The black line overlayed onto the histograms in Fig.~\ref{fig:CubeHistos} is the reconstructed mass distribution, or density of states, of the cube obtained using MBAR according to Eq.~(\ref{eq:MBARHistos}).
Notice that, as expected, the region of large $r$ leading to $\sqrt{d}a = 5$ contains very little volume.

\begin{figure}
    \centering
    \includegraphics[width=0.96\columnwidth]{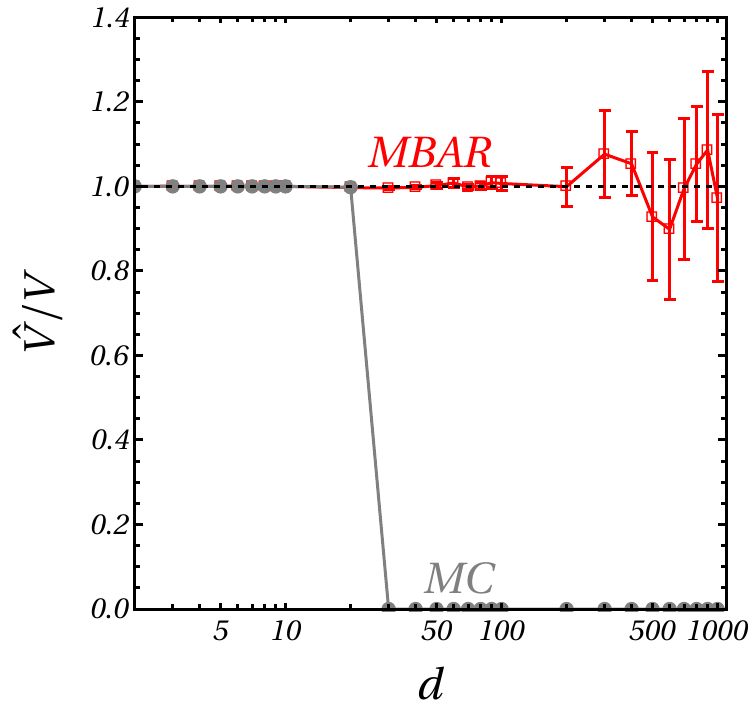}
    \caption{\textbf{Volume estimates.} 
    Ratio of measured volume $\hat{V}$ to expected volume $V$ of the unit cube against the dimensionality $d$, using brute-force Monte Carlo with $N_s = 10^8$ points in the smallest ball containing the cube (gray disks) and MBAR (red squares) with $K = max(64,d/5)$ random walks linearly spaced in $r$, each with $N_s = 5\times10^5$ proposed steps.
    Error bars are $95\%$ confidence intervals for the mean, using statistics across $10$ runs.
    }
    \label{fig:CubeVolumes}
\end{figure}

We use the same procedure when varying the dimensionality of the cube.
The result for the ratio $\hat{V}/V$ between the measured volume and the real volume of the cube, obtained using the fastMBAR~\cite{Ding2019} implementations of MBAR, is shown in Fig.~\ref{fig:CubeVolumes} as a function of the dimensionality (red squares).
The results are compared with a brute-force Monte Carlo method using the same order of magnitude of total number of samples, $N_s = 10^8$ (gray disks).
Notice that the $x$ axis is in log scale.

In spite of the finite number of samples, and of the rather crude choice of $K_d$, the error of the MBAR measurement grows roughly linearly with $d$ and remains below $10\%$ even as $d$ reaches $500$, while the Monte Carlo estimate essentially measures zero volume when dimensionality reaches $d=30$, the point at which the ratio of the volume of the cube to that of the smallest circumscribed ball is roughly $1/N_s$.
Notice that in $d=500$, that ratio is of the order of $10^{-156}$, meaning that na\"{i}ve Monte Carlo would be absolutely unfeasible.
Here, using our approach instead of na\"{i}ve MC sampling, \textit{with comparable numbers of samples}, increases the maximum dimension for which a volume can be measured, albeit with some statistical error, from 30 to thousands of dimensions.
Since the number of samples used per walk in this example is still rather modest ($5 \times 10^5$), the statistical error can be reduced by making the random walks longer.
\begin{figure}
    \centering
    \includegraphics[width=0.96\columnwidth]{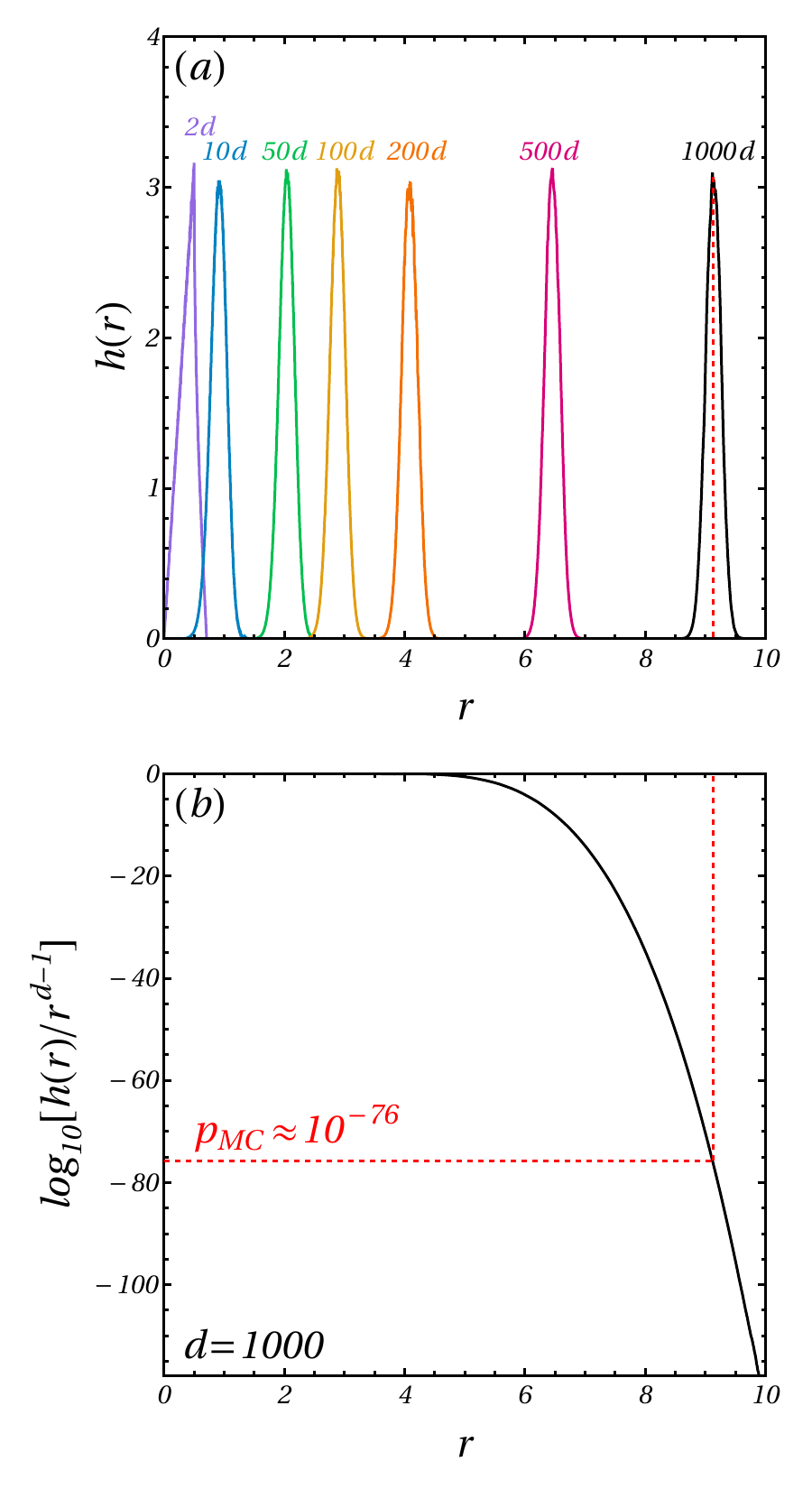}
    \caption{\textbf{Mass distribution estimate.} 
    $(a)$ Mass distribution of the unit-sidelength hypercube, $a = 0.5$, as estimated from MBAR, using the same runs as in Fig.~\ref{fig:CubeVolumes}, in $d = 2, 10, 50, 100, 200, 500, 1000$  dimensions.
    $(b)$ Decimal log of the ratio between the mass distribution and that of the smallest ball containing the whole cube for $d = 1000$.
    The dashed red lines indicates the location $\sqrt{d/3}a$ of the maximum for $d =1000$, and the value of the ratio it corresponds to in $(b)$.
    We indicate the probability $p_{MC}$ of landing a single point inside the cube when drawing points uniformly from that specific hyperspherical shell.
    }
    \label{fig:CubeDOS}
\end{figure}
Our approach also yields the mass distribution, as reconstructed by unbiasing the histograms of individual random walks, according to Eq.~(\ref{eq:MBARHistos}).
In Fig.~\ref{fig:CubeDOS}$a$ we show the hypercube mass distribution, $h(r)$, for a few dimensionalities up to $d = 1000$.
As expected, as $d$ increases, $h(r)$ tends to a sharp peak centered at $\sqrt{d/3}a$, indicated by a red dashed line for $d = 1000$.

To illustrate the difficulty of measuring not just the volume, but also the mass distribution of the cube in such high dimensionality, in Fig.~\ref{fig:CubeDOS}$b$ we plot $h(r)/r^{d-1}$ for $d=1000$, which amounts to the probability of landing inside the cube when sampling uniformly on the surface of the hypersphere with radius $r$.
Because the intersection between the cube and a sphere becomes very small, we plot the base-$10$ log of that ratio.
A value of particular interest is that obtained for the mode of $h(r)$, as it gives an estimate of how many trials would be needed to find a single point inside the cube if we were performing brute-force Monte Carlo, restricted to the shell of the cube that contributes most to its mass.
In $d=1000$, we find that this number is of the order of $10^{76}$. Even if one could somehow guess the radius of the shell that contributes most to a high-dimensional volume, direct Monte Carlo sampling of that shell alone is in general impossible.
In fact, we manage to reconstruct the density of states reliably until points that would require $10^{120}$ Monte Carlo shots on a sphere, using only $10^{8}$ points across all walks.

Altogether, we have shown for a simple example that using importance sampling and a free energy estimation method (here, MBAR) enables us to measure high-dimensional volumes much more efficiently than brute-force Monte Carlo by virtue of the correlations between successive positions of the random walkers, and by avoiding getting stuck in singular features of the domain thanks to parallel tempering.

Going beyond the simple example given here, the contrast in efficiency can be made even starker by refining the precise design of the algorithm we presented.
First, one could choose a different set of biasing potentials than the one used above, which reproduces the strategy of Ref.~\cite{Martiniani2017a}.
For instance, the scaling we used for the number of replicas is likely to be a worst-case scenario, as it assumes equally-spaced distributions in $r$ that are all as narrow as the narrowest one, $h_{k_{max}}$.
One can instead use an iterative choice for the values of $k$'s such that the mode of $h_{k_{n+1}}$, $r^*_{n+1}$, lies at $r^*_n + \sigma_n$ with $\sigma_n$ the standard deviation of $h_{k_n}$.
Assuming perfect Gaussian distributions for each random walk, and requiring that the values go up to the mode of the mass distribution of the cube minus the standard deviation, $\sqrt{d/3 - 1/15} - \sqrt{4/(45d)}$, 
some simple algebra leads to the asymptotic scaling $K_d \sim d^{1/2}$.
Since MBAR does not require the biasing potentials to be harmonic, it is likely that this scaling could be brought down even further with an altogether different choice of biasing potentials, e.g. by choosing a series of potentials of the form $U_i(r) = r^{1-d}\exp[-k(r-r_i)^2]$, where $r=|\mathbf{r}-\mathbf{r}_0|$, $k$ is a fixed width, and $r_i$ is a tunable scalar distance from $\mathbf{r}_0$.

Second, one could accelerate the diffusion of the random walks by ``cloud sampling'', which amounts to biasing the Monte Carlo sampling on the average weight of a larger number of trial points at each step~\cite{Martiniani2017}.
Alternatively, one could introduce some deterministic drift in the random walks, so as to make them closer to recently-proposed piecewise-deterministic processes~\cite{Chevallier2022}, or so-called Galilean Monte Carlo~\cite{Skilling2012,Griffiths2019} methods, where particles travel following straight lines and bounce on walls instead of diffusing around.

The higher exploration efficiency afforded to us by this method is far from being free in general basin volume computations: in order for the random walk to remain within the basin, we need to query an “oracle” at every step to determine if we are inside or outside the shape of interest~\cite{Frenkel2017}.
While this oracle is a simple geometric condition in the case of the hypercube, for basins of attraction in a many-body energy landscape we must solve for the path of steepest descent at every step of the random walk~\cite{Wales2003}, meaning that a single Monte Carlo step typically requires hundreds of energy (function) evaluations.
In liquids, for instance, the cost of one such energy evaluation scales at best like the number of particles, $N$, which needs to be large for thermodynamic properties to be measured accurately.
While costly, we will show in Sec.~\ref{sec:granularentropy} that this cost is manageable in a practical example.

\subsection{Unbiasing sampled volumes}

Recall that our original aim was to estimate the mean basin volume, $\langle v \rangle$, in Eq.~\ref{eq:mbv}.
So, after having measured the volume of a large number of basins of attraction, we must deal with the issue that a naïve average over the basin volumes of randomly sampled energy minima is in general biased.
Indeed, if one samples configuration space uniformly and tags each point with its basin of attraction, each basin is sampled with probability $p_i = v_i / \mathcal{V}$, that is to say proportionally to its volume.
Thus, to compute a Boltzmann entropy $S_B = \log \Omega + const.$, we need to perform a fit of the observed basin volume distribution using a putative functional form to undo the bias and obtain the true mean, $\langle v \rangle$.
Concretely, if the measured (biased) distribution of volumes is called $B(v)$, and the unbiased (true) distribution $U(v)$, one can write
\begin{align}
    B(v) = \mathcal{N} U(v) v,
\end{align}
where the $v$ is the bias proportional to volumes, and $\mathcal{N}$ is a normalisation factor.
By integration, one finds that the unbiased mean is given by
\begin{align}
   \mathcal{N}^{-1} = \langle v \rangle = \left[\int\limits_{0}^{\mathcal{V}} \frac{\mathrm{d}v}{v}B(v) \right]^{-1},
\end{align}
where one typically needs to fit $B(v)$ to some functional form, for instance a parametric generalized Gaussian distributions, or a nonparametric kernel density estimate.

The Shannon entropy, $S_S = - \sum_{i=1}^\Omega p_i \log v_i + const.$, instead is by definition the \textit{biased} average of the negative log-volumes of basins, and it can be obtained directly with no assumptions.

All in all, given a proper algorithmic basis, basin-volume measurements provide a generic way of computing the Gibbs-Shannon entropy of any dynamical system whose properties are controlled by the ensemble of its steady-state structures.
In the following, we shall first show how this idea was applied to the special case of granular packings, and then how it could answer open questions in fundamental physics.

\section{The case of granular entropy\label{sec:granularentropy}}
\begin{figure}
\begin{center}
\includegraphics[width=0.475\columnwidth]{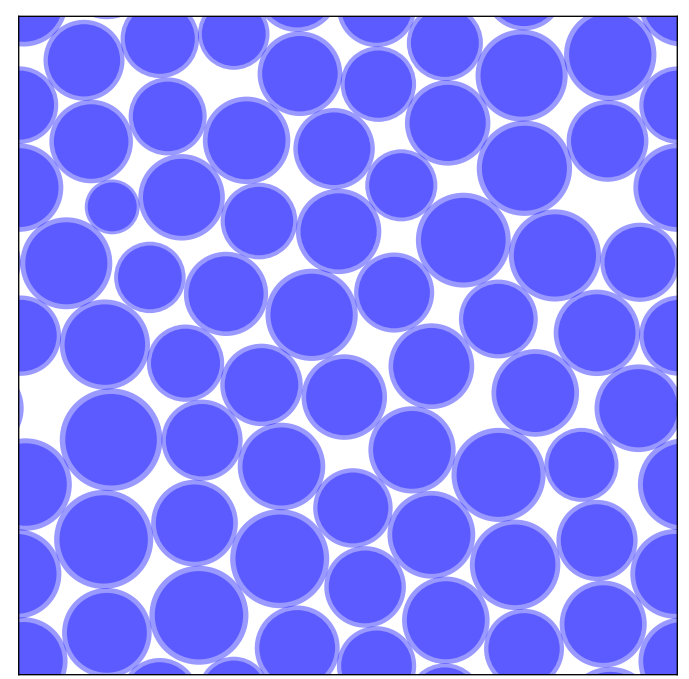}
\includegraphics[width=0.46\columnwidth]{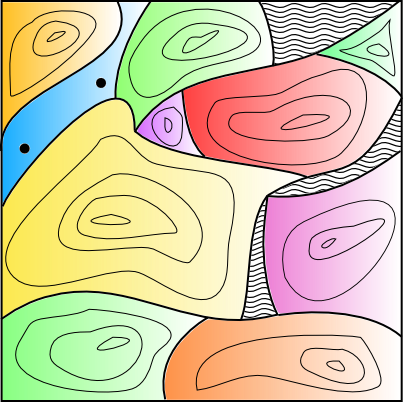}
\end{center}
\caption{\textbf{Energy landscape of hard-WCA particles.}
Left: Snapshot of jammed packing of polydisperse disks with hard cores (dark shaded) plus soft repulsive coronas (light shaded).
Right:  Illustration of configurational space for jammed packings. The hatched regions are inaccessible due to hard-core overlaps. 
Single-colored regions with contour lines represent the basins of attraction of distinct minima. Blue region with solid dots indicates the coexisting unjammed fluid region (observed only for finite size systems) and hypothetical marginally stable packings.
Figures adapted from Ref.~\cite{Martiniani2017a}.
} \label{fig:packingIllu}
\end{figure}
\begin{figure*}
\begin{center}
\includegraphics[width=1.9\columnwidth]{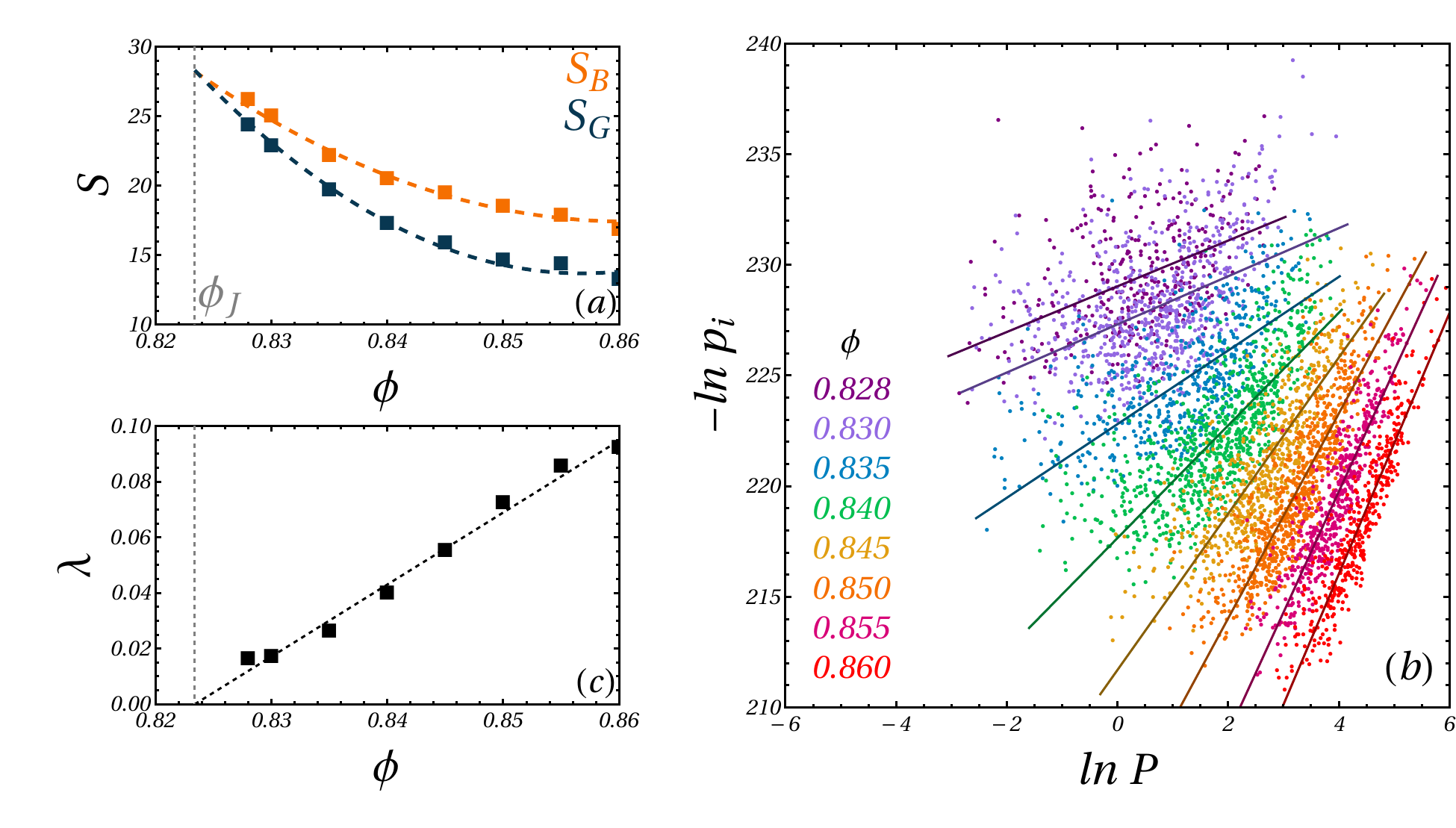}
\end{center}
\caption{\textbf{Checking Edwards' hypothesis.}
$(a)$ Gibbs-Shannon entropy $S_G$ (blue) and Boltzmann entropy $S_B$ (orange) as a function of $\phi$, as obtained from a parameter-free kernel-density estimate (KDE) fit of the distribution of basin volumes. 
The dashed curves are second-order polynomial fits.
As $\phi$ approaches the unjamming density for $N=64$ particles $\phi_J = 0.82$ (dashed gray line), $S_G$ tends to $S_B$, implying equiprobability of all states precisely at unjamming.
$(b)$ Scatter plot of the log of the probability of landing in a given basin, $p_i$ (proportional to the volume $v_i$) against the log of the pressure $P$ measured in that basin. Power-law scaling relations (solid lines) are found for several densities indicated in the inset.
$(c)$ Exponents found for the power laws of $(b)$, plotted against $\phi$.
The dashed black line is a linear fit, and the dashed gray line indicates $\phi_J$.
The raw data from Ref.~\cite{Martiniani2017a} was used here and re-analyzed independently.
} \label{fig:granularresults}
\end{figure*}
The first application of basin-volume calculations to measure an entropy was performed in the context of granular packings in two dimensions~\cite{Xu2011, Asenjo2014, Martiniani2016, Martiniani2017a}.
Due to their athermal nature, granular systems cannot be described by Gibbsian statistical mechanics~\cite{Gibbs1902}.
Nevertheless, in the late 1980s, Edwards and Oakeshott~\cite{Edwards1989} proposed that the collection of stable packings of a fixed number of particles in a fixed volume could play the role of an ensemble, and that one could arrive at a statistical mechanical formalism by making the assumption that all stable packings are equally probable once the system has settled in a jammed state. 
In other words, in the Edwards’ ensemble, jammed states occur with uniform probability measure (also known as Edwards’ measure~\cite{Baule2018}). 
The existence of a Boltzmann-like distribution in volume and stress in granular media has been supported by a number of experiments~\cite{Aste2006, Aste2008, McNamara2009, Puckett2013, Zhao2014, Bililign2019, Sun2020, Yuan2021}, lending credence to Edwards’ theory.
However, theoretical checks of Edwards' key hypothesis on the equiprobability of packings was largely considered as being not directly testable in simulations, as it seemingly required an explicit enumeration of all possible packings.

This problem can be rephrased as a basin volume measurement: consider a system of polydisperse disks with a hard core and a soft corona, modelled as a shifted Weeks-Chandler-Anderson (WCA)~\cite{Weeks1971} soft potential (see Fig.~\ref{fig:packingIllu}).
Consider $N$ such particles, put them into a periodic box with sidelength $L$ at a packing fraction $\phi$.
The system is endowed with a many-body energy landscape that, at high enough densities, contains many different basins.
At the bottom of each basin lies a single local minimum, that can be interpreted as a zero-temperature configuration of the system. 
At the characteristic unjamming density $\phi_J$ (the point at which the system goes from a liquid to a disordered solid), each minimum corresponds to a packing of non-frictional hard particles.
Among these minima, there is a subset of mechanically stable configurations.
Depending on the system of interest, it is common to define mechanical stability through a non-zero bulk modulus, which is equivalent to saying that the number of contacts verifies $N_c \geq d (N_{nr} -1)$, where $N_{nr}$ is the number of particles that are not \textit{rattlers} (mobile particles)~\cite{Goodrich2014}, thus defining the ensemble of collectively jammed states~\cite{Atkinson2013}.
A smaller subset of minima is that of states stable not only against compression, but also shear: the condition of non-zero shear modulus this time imposes $N_c \geq d(N_{nr} - 1) + 1$, which defines the ensemble of \emph{strictly} jammed states \cite{Atkinson2013}.
In practice, we restrict ourselves to the latter case \cite{Martiniani2017a}, which is a more broadly accepted definition of jamming.
Therefore, by implementing the machinery of Sec.~\ref{sec:basinvolume} to the energy landscape of hard-WCA particles, and restricting the set of relevant basins to mechanically stable configurations, one can estimate the Boltzmann and Gibbs-Shannon entropy of granular packings.

The main result, originally obtained in Ref.~\cite{Martiniani2017a} for $N = 64$ hard-WCA disks, is reproduced in Fig.~\ref{fig:granularresults}$a$.
This plot shows that the Gibbs-Shannon entropy, $S_G = - \sum_{i=1}^\Omega p_i \log p_i + const.$, and the Boltzmann entropy $S_B = \log \Omega + const.$, both evaluated using basin volume measurements, converge precisely at $\phi_J \approx 0.82$, implying that only at this density is Edwards' hypothesis, $p_i = 1/\Omega$, verified.
It can be checked through finite-size analysis~\cite{Martiniani2017a} that this packing fraction is consistent with the lower range of values for which the system unjams at $N=64$ following the same preparation protocol, so that the two entropies coincide only at unjamming.
Note that, while the value of the unjamming density observed in compression experiments depends on the precise preparation protocol, what we report is a property of the energy landscape of jammed packings, which is a generic feature of the system independent of preparation recipes.

Furthermore, as shown in Fig.~\ref{fig:granularresults}$b$, the authors also found that above jamming, there exists a robust power law relationship, $p_i \sim P^{-N \lambda(\phi)}$, between the probability of observing a packing and its pressure.
This power law suggests a hierarchical structure of the energy landscape of hard-WCA particles, where low-energy minima have large basin volumes and high-energy minima have small volumes, see Fig.~\ref{fig:sketchpowerlaw} for an illustration of this property.

Finally, as shown in Fig.~\ref{fig:granularresults}$c$, the exponent $\lambda$ decreases roughly linearly as the packing fraction approaches jamming from above, and reaches $0$ at $\phi_J$.
In other words, at jamming, the volume of basins becomes a flat distribution that does not depend on the pressure in the system, adding to the evidence that all basins are equiprobable at $\phi_J$.
As shown in App.~\ref{app:powerlaw}, this argument on $\lambda$ can be made more formal, and the difference between the Boltzmann and Gibbs entropies can be written as
\begin{align}
   \lim_{N\to\infty}\frac{1}{N}\left[S_B(V) - S_G(V)\right] = O(\lambda^2).
\end{align}
In other words, in order for the two entropies to agree at a given density in the thermodynamic limit, the exponent $\lambda$ \textit{must} go to $0$ as observed.

All in all, a first application to granular packings has shown the potential of the method proposed in Sec.~\ref{sec:basinvolume}.
By replacing an intractable enumeration problem by a simpler sampling problem, the basin-volume method allowed to test an hypothesis that had been left unchecked for 30 years.
It also yielded insight into new physics at the jamming transition, by revealing a hierarchical structure of the basins of attraction above jamming.

\section{The road ahead: from packings to generic dynamical systems\label{sec:basinseverywhere}}

In the rest of the paper, we propose perspectives for the basin-volume method, ranging from extensions of the problem of granular packings to completely unrelated problems.

\subsection{Future work on granular packings}

As discussed in Sec.~\ref{sec:granularentropy}, the basin-volume method led to a direct observation that Edwards' hypothesis is only valid strictly at jamming in $d=2$.
There are however still several aspects of this problem that remain unchecked, and could be addressed by the very same method.

First, these results were obtained within the isochoric ensemble, approaching unjamming where the measured pressure vanishes, $P \to 0^{+}$.
It has been argued that the equiprobability of packings could break down when switching to the isobaric ensemble~\cite{Gao2006}, \textit{viz.}, when allowing volume fraction fluctuations through particle inflation and deflation to maintain constant hydrostatic pressure, or more generally in the isostress ensemble that also constrains the shear stresses.

The isobaric and isostress ensembles can be explored within a basin volume framework. To do so, in the spirit of Parrinello-Rahman barostats~\cite{Parrinello1980,Parrinello1981,Parrinello1982}, the trick is to allow for deformations of the simulation box (both isotropic compressions and constant-volume deformations), and to replace the \textit{energy} landscape, $E$, by an \textit{enthalpy-like} landscape, $H$~\cite{Goodrich2014}, where 
\begin{align}
H = E - \overline{\overline{\sigma}} : \overline{\overline{\varepsilon}} V_0
\label{eq:enthalpy}
\end{align}
explicitly contains the dependence on the full stress and strain tensors $\overline{\overline{\sigma}}$ and $\overline{\overline{\varepsilon}}$ through their Frobenius inner product $\overline{\overline{A}}:\overline{\overline{B}} = \sum_{i,j} A_{ij} B_{ij}$, as well as a reference volume $V_0$ of the simulation box.
Note that a subcase of this strategy is that in which only isotropic compression is allowed,  $\overline{\overline{\varepsilon}} = V/V_0 \overline{\overline{I}}$, with $\overline{\overline{I}}$ the identity tensor.
In that case, only the trace $Tr[\overline{\overline{\sigma}}] \equiv - P$  participates to the box deformation term, leading to the usual definition of enthalpy with respect to the hydrostatic pressure $P$, namely $H = E + P V$.
Basins are then defined through steepest descent paths that lead to the same enthalpy minimum, in a configuration space comprising not only particle positions, but also $d(d+1)/2$ parameters describing simulation box deformations \cite{Goodrich2014}.
In practice, one would impose a finite but small pressure, and look at the asymptotic results as $P\to 0^+$.
Indeed, a technical difficulty in this approach is that, at exactly zero pressure, fluid solutions with arbitrarily large volumes trivially minimize the enthalpy.
One would then need to constrain the maximal volume of the system to avoid converging exclusively to fluid states. 

Finally, to perform basin volume calculations in the enthalpy landscape, a modification of the Monte Carlo algorithm used for the volume estimate is required.
In the isochoric case presented above, the volumes of basins are computed by sampling configurations with constant $N$ and $V$, using an oracle defined through a minimum of the energy, and umbrella sampling on the particles' positional degrees of freedom.
In the isobaric case, this time, one needs to sample configurations with constant $N$ and $P$, with an oracle defined through a minimum of the enthalpy, and using umbrella sampling not only on the particles' positions, but also on the degrees of freedom of the box shape.
This requires the inclusion of independent shearing and stretching (viz., box deformations) moves in our Monte Carlo sampling~\cite{Najafabadi1983,Yashonath1985,Baldock2016, Baldock2017} as the building block of the random walks.
Each of these moves is accepted or rejected according to a Monte Carlo criterion set by a combination of the oracle, which checks that the new point still falls to the same minimum of the enthalpy, Eq.~\ref{eq:enthalpy}, and the umbrella sampling biasing potentials that contain box deformations.
Implementing this algorithm as part of a basin volume calculation, while numerically costlier than the usual isochoric strategy, would constitute an important check on the validity of the Edwards hypothesis.

Second, one may reasonably wonder whether the results are still observed in higher dimensions of space, notably in $d=3$, where many open questions remain to be answered to fully understand the ensemble of jammed states, thus constituting a topic of current research~\cite{Anzivino2022}.

\begin{figure}
    \centering
    \includegraphics[width=0.96\columnwidth]{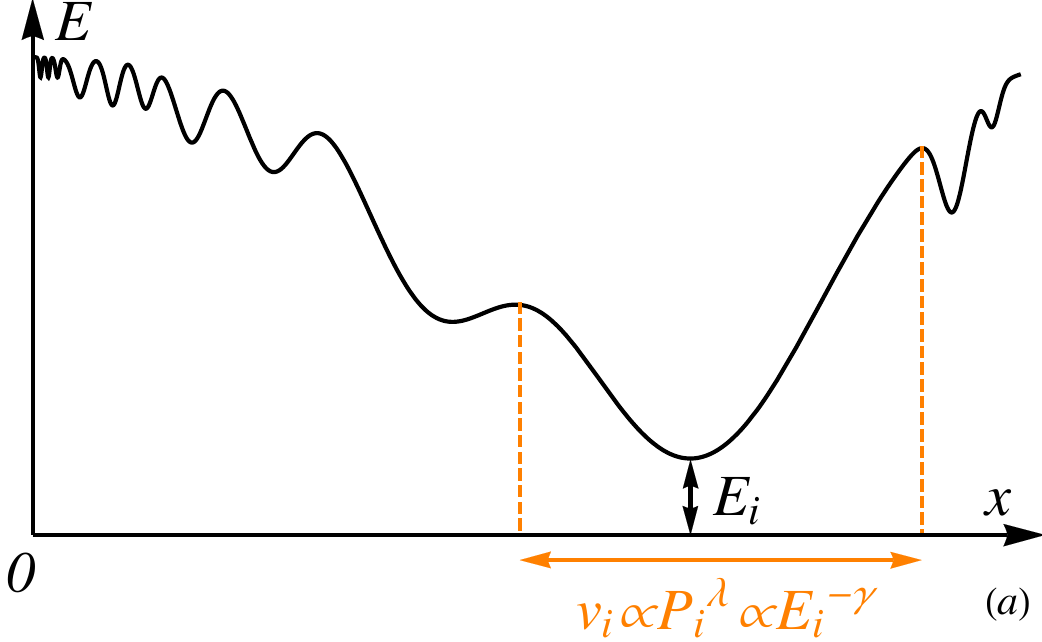}\\
    \vspace{2mm}
    \includegraphics[height=0.47\columnwidth]{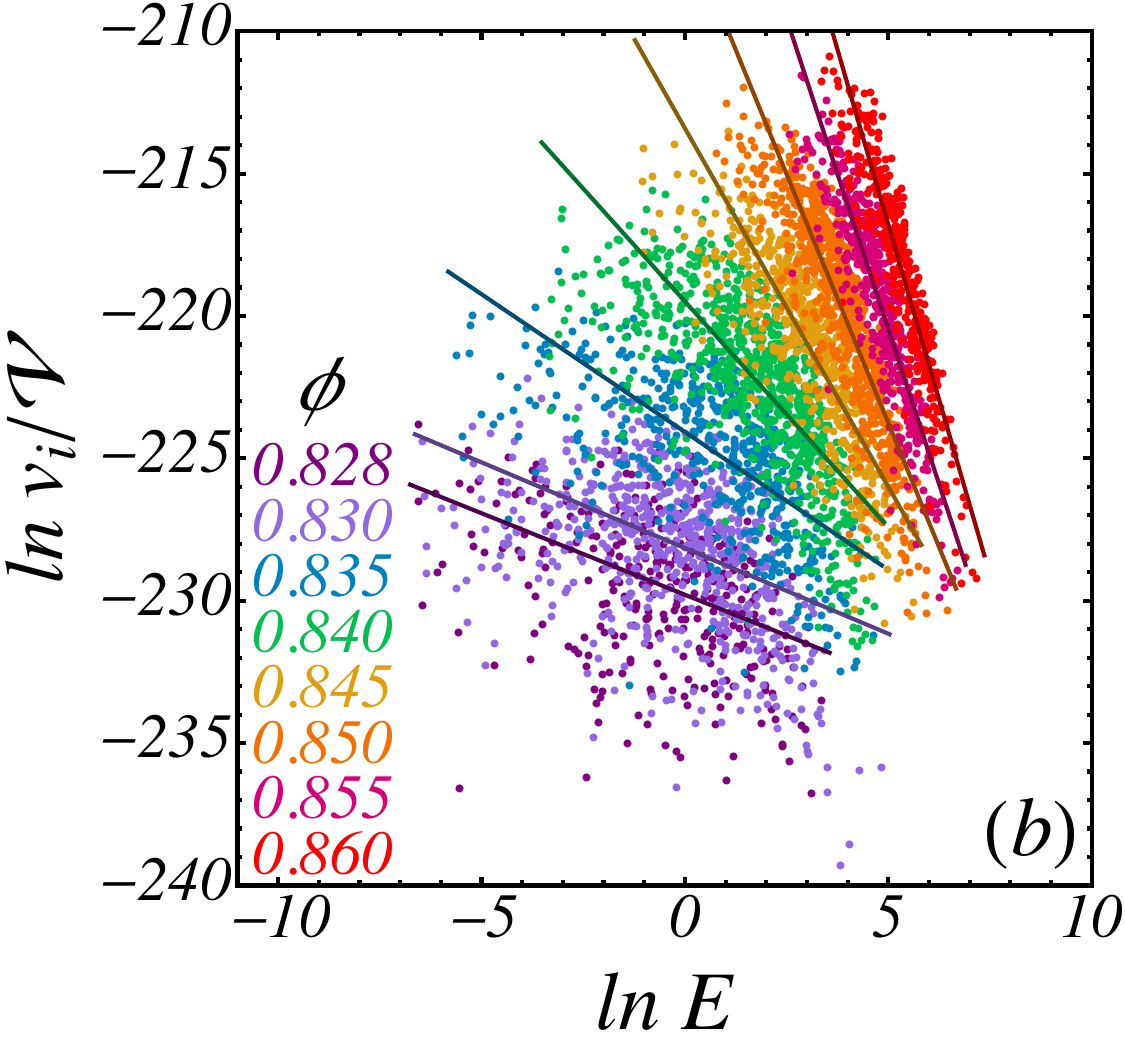}
    \includegraphics[height=0.47\columnwidth]{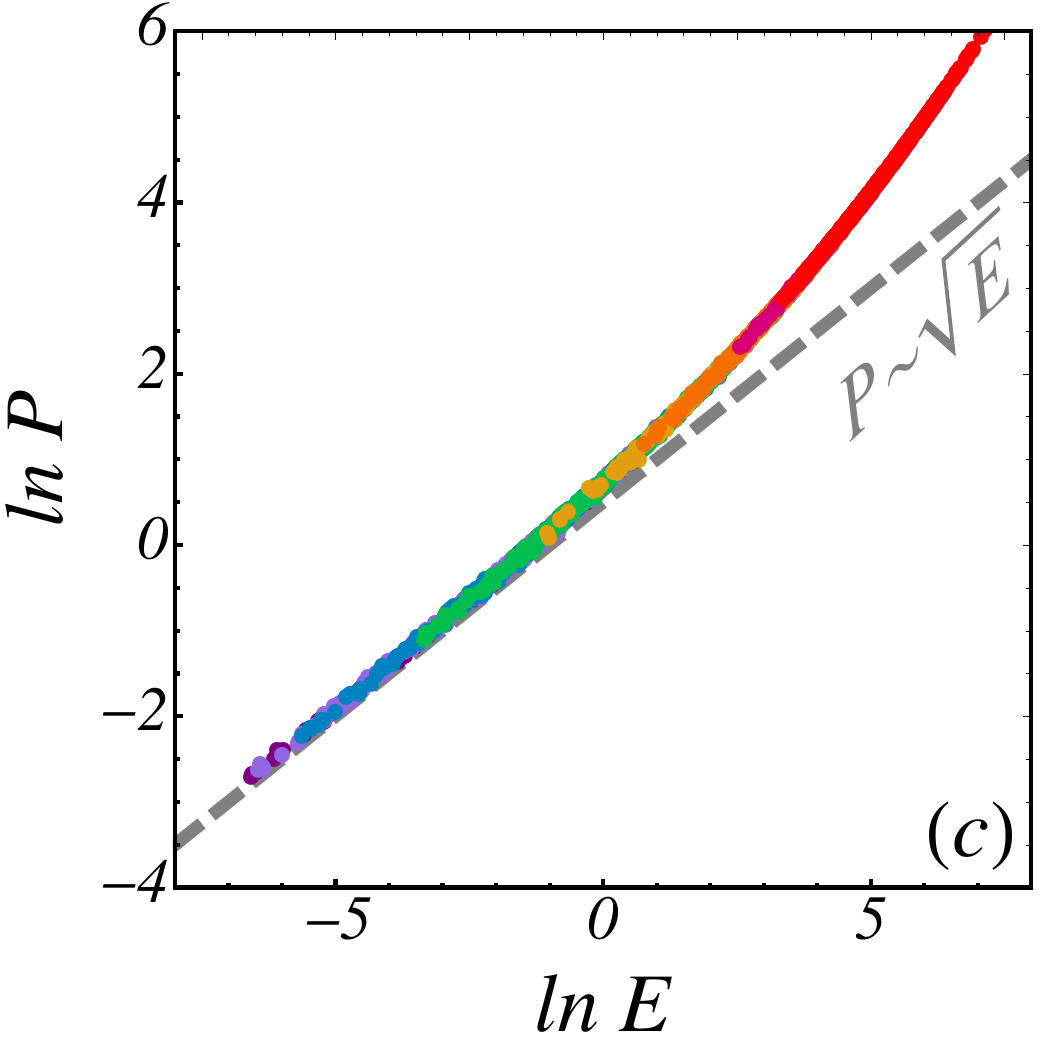}
    \caption{\textbf{Hierarchical energy landscape of soft sphere packings.}
    $(a)$ Sketch: the power law between basin volume, $v_i$, and pressure suggests that a minimum lying at $E_i > 0$ has a basin with volume $v_i \propto E_i^{-\gamma}$ with $\gamma$ a positive exponent.
    We verify this scaling relationship in $(b)$, using the same data as in Fig.~\ref{fig:granularresults}.
    Panel $(c)$ shows a smooth dependence of $P$ as a function of $E$, with an asymptote $P\propto\sqrt{E}$ at unjamming.
    }
    \label{fig:sketchpowerlaw}
\end{figure}

Finally, the observation that, in the isochoric ensemble and above jamming the basin volumes are linked to the pressure of the packings via a power-law suggest that configuration space is tiled by a hierarchy of basins (see Fig.~\ref{fig:sketchpowerlaw}$(a)$), which has neither been fully appreciated, nor explored.
We check that such a hierarchical dependence exists in Fig.~\ref{fig:sketchpowerlaw}$(b)$, using the same data as in Fig.~\ref{fig:granularresults}, and report a density-dependent power-law between the volume $v_i$ of a basin and the energy at the corresponding minimum.
This power law results from $P$ being a smooth function of the energy, as shown in Fig.~\ref{fig:sketchpowerlaw}$(c)$, where we emphasize that unjamming is accompanied by an asymptote $P\propto \sqrt{E}$.
This behaviour is expected near unjamming, where overlaps are small so that the pair potential is well captured by a harmonic approximation, $U(r)\sim(r-r_0)^2$ \cite{Martiniani2017}, and the pressure contribution from one interacting pair, as obtained from a virial expression, reads $p \sim r_0 (r - r_0) \sim \sqrt{U}$ which, summed over, can be shown to yield $P\sim \sqrt{E}$ as $E\to 0$.

\subsection{Beyond granular jams: glassy systems}

Based on our discussion thus far, a reader may think that basin volume calculations can only be used to measure granular entropies.
In fact, these methods could be used to measure a number of indicators for the shapes of the basins, in addition to the radial mass distribution, as well as their neighborhoods.
Such measurements could have deep fundamental implications, if one does not consider a granular (or zero-temperature) system, but a jammed liquid at a finite temperature.
Indeed, in a finite-temperature glassy system, the shape of basins controls transition rates and relaxation times~\cite{Castellani2005, Ros2019,Rizzo2021, Folena2022}, which has prompted a large body of recent work on the enumeration and characterization of minima of rough landscapes~\cite{Ros2019,Ros2019b,Ros2021,KentDobias2021,LacroixAChezToine2022}.

Achieving a clear picture of the basins of attraction in the configuration space of a dense liquid in physical dimensions would therefore be very exciting, as it would open avenues for fundamental checks regarding glassy dynamics.
In an exact mean-field treatment of hard sphere glasses, a Kauzmann transition known as random first-order transition (RFOT) is attained~\cite{Charbonneau2017, Franz1995, Parisi2010}.
In this framework the configurational entropy $S_{\mathrm{conf}} = \log \Omega_F$ is defined in terms of the number of free energy minima $\Omega_F$, so that the notion and number of “glassy states” is well defined and the ideal glass transition results from the population of glassy states becoming subextensive in system size.
In finite dimensions, free energy minima are not infinitely long-lived and the precise definition of a “glassy state” remains debated~\cite{Berthier2019}.
A popular take on the problem was proposed by Stillinger and Weber~\cite{Stillinger2015, Stillinger1982, Stillinger1995}, who suggested that the supercooled glassy states would reside close to the minima of the potential energy surface, known as inherent structures (IS) of the liquid.
For each IS there is an associated basin of attraction, that is the set of all initial conditions leading to the $i$-th IS by steepest descent. 
The partition function can then be expressed classically as a sum over the individual IS, $Q = \sum_{i=1}^\Omega q_i$, where, for $\beta = 1/k_BT$,
\begin{align}
    q_i = \int_{\Gamma_i} e^{-\beta \mathcal{U}(\bm{r})} \mathrm{d}\bm{r},
\end{align}
and $\mathcal{U}(\bm{r})$ is the interaction potential.
Until recently, $q_i$ could not be computed, and the accepted operational definition for the configurational entropy was $S_{\mathrm{conf}} = S_{\mathrm{liq}} - S_{\mathrm{harm}}$, where $S_{\mathrm{liq}}$ is the total entropy computed by thermodynamic integration, and $S_{\mathrm{harm}}$ is the vibrational entropy computed from the harmonic approximation (with or without anharmonic corrections), averaged over many IS~\cite{Sciortino2005}.
Using this and another approach based on a Frenkel-Ladd-like computation, Berthier and coworkers recently produced compelling results suggesting that the change in $S_{\mathrm{conf}}$ for supercooled liquids in $d=3$ is consistent with the existence of an ideal glass transition at $T_K > 0$~\cite{Berthier2019,Berthier2017,Ozawa2018, Nishikawa2022}.

Through basin volume calculations, it is possible to compute $q_i$ directly, and thereby obtain a general approach to estimating the configurational entropy of supercooled liquids, without resorting to any approximations. In particular, this method does not need to assume large system sizes, and could allow to precisely study finite-size effects in complex free energy landscapes, a problem that is nontrivial even in simple theories in $1d$~\cite{Rulquin2016}.

Finally, in the context of glassy systems, free energy methods could in principle be used to obtain not just the (Boltzmann weighted) volume of the basins of attraction of inherent structures, but also more complicated information about their geometry, topology, and connectivity (e.g., what is the chromatic number for the tiling of basins in the energy landscape?~\cite{frenkelpriv}).
Understanding the properties of the most probable paths between minima of the energy landscape is of particular interest to understand the dynamics of relaxation in glasses, and could help bridge the gap between microscopic dynamics and mesoscopic models that recently shed new light on the interpretation of the relaxation spectrum of glassy materials~\cite{Scalliet2021, Guiselin2022}.

\subsection{Generic dynamical systems: ecosystems, neural networks, and more.}

\begin{figure}
\begin{center}
\includegraphics[width=0.92\columnwidth]{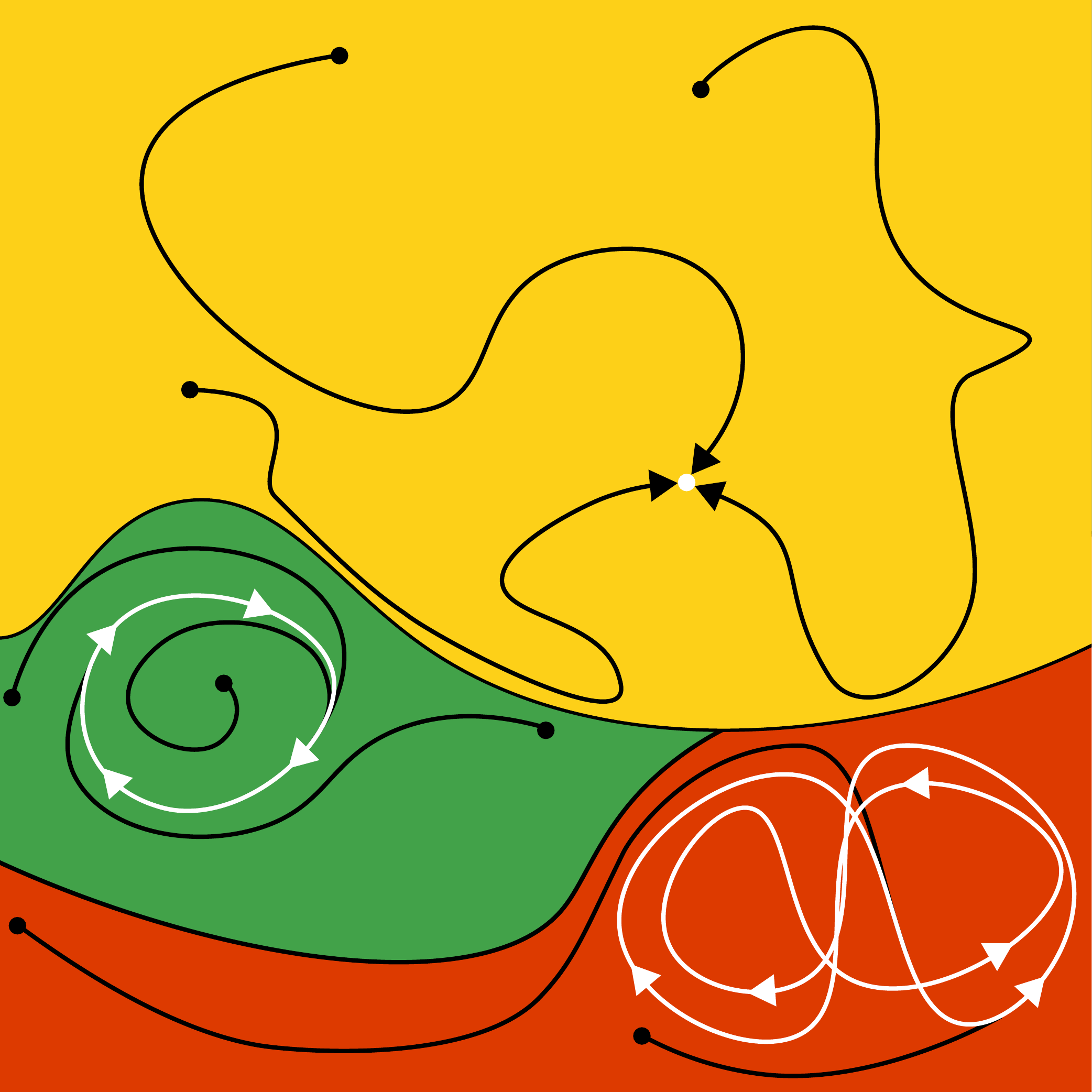}
\end{center}
\caption{\textbf{Generalised basins for dynamical systems.}
In a generic dynamical system, basin of attractions can be associated not only to fixed points, but also limit cycles, or any kind of stable attractor of the dynamics.
Black arrows represent example trajectories, that fall into steady-state structures (white).
These structures can be fixed points (yellow basin), simple limit cycles (green basins) or any complicated high-dimensional attractor (red basin).
} \label{fig:generalisedbasins}
\end{figure}

In the remainder of this paper, we discuss what we believe to be some of the most exciting opportunities for basin-volume methods, namely their applications to generic dynamical systems. Indeed, going back to the method presented in Sec.~\ref{sec:basinvolume}, we never explicitly relied on the existence of an underlying energy function.
In principle, one can define any dynamics of their choice, let them evolve, and classify initial conditions according to which steady-state or dynamical attractor they eventually fall into, as sketched in Fig.~\ref{fig:generalisedbasins}.
One can then always define the Shannon entropy associated with the volumes of these attractors, and use it as a generic descriptor of the system for a given set of parameters.
This approach is completely general and can be applied to any system with regular enough dynamics that they admit attractors. We propose a few promising leads for such approaches.

There are many examples of dynamical systems where such an approach would be relevant.
For instance, mapping basins of attraction is an ubiquitous problem in control theory, where approximate methods for measuring basins have been proposed to determine the stability of the flight control-law of the F/A-18 Hornet aircraft with respect to large perturbations~\cite{Chakraborty2011}.
Enumerating the number of stationary points, and their distribution, for certain classes of random functions, is a classical problem in mathematics and statistics~\cite{Kac1943, Farahmand1986, Bogomolny1992, Edelman1995, Rojas1996, Kostlan2002, Fyodorov2004, Malajovich2004, Azais2005, Armentano2009, Fyodorov2012, Fyodorov2012a, Fyodorov2013, Fyodorov2015, Cheng2015}.
In particular, in combinatorial optimization, the size and connectedness of the space of solutions that satisfy a large number of constraints controls how easy it is to find a solution, which has implications in a variety of real-life problems ranging from computer science to optimal transport~\cite{Krzakala2007,Krzakala2008,Zdeborova2008,TayaraniNarajan2013}.
Understanding the structure of high-dimensional basins of attraction is also of great theoretical interest, as the problem of mapping high-dimensional basin volumes has been described as uncharted mathematical territory~\cite{Zhang2021}.
It is also a current challenge in cosmology, where characterising the number and volume of basins of attraction in axion landscapes constitutes a promising lead for elucidating the cosmological constant problem~\cite{Bachlechner2019}.

A particular example of dynamical systems where basin volume methods could prove fruitful is the study of ecosystems, i.e. ensembles of populations following Lotka-Volterra dynamics~\cite{Lotka1910}.
In the limit of symmetric antagonistic interactions, such systems have been shown to map onto glassy Hamiltonian systems, which leads to aging dynamics in a rugged landscape, the properties of which have been studied analytically at the mean-field scale~\cite{Bunin2017, Biroli2018, Roy2019, Altieri2021}.
Therefore, in this setting, one can apply the strategy of basin volume calculations to determine the existence and nature of phase transitions in the limit of a finite number of species.
Furthermore, since an Hamiltonian structure is not needed to define an entropy via basin-volume calculations, the same approach can also be used in the much more complicated case of non-reciprocal, not-all-antagonistic interactions.
In this scenario, qualitative results, both experimental and theoretical, have shown that stable structures tend to emerge in steady state~\cite{Barbier2018,Bunin2018,Barbier2021}.
Basin-volume approaches could provide insight into these systems, for instance by quantifying how many ecosystems are possible in steady-state for a given mix of interactions, or which ecosystem is the most typical.

Finally, another exciting lead is the application of basin-volume methods to Deep Neural Networks (DNNs) \cite{ballard2017energy}. It has been widely observed that the parameters learned by DNN during training correspond to ``flat minima'' of the loss-function landscape~\cite{Hochreiter1994, Hochreiter1997, Baldassi2020, Feng2021, Zhang2021a, Pittorino2021}.
Basin volume calculations may enable researchers to answer questions concerning the structure of the error landscapes of DNNs and to identify the relationship between the probability of finding a given solution, its flatness and its generalization performance. 
Addressing these questions would have a significant impact on our understanding of generalization in deep learning systems with implications for high-stakes applications such as transportation, security and medicine.

\section{Conclusions}
In this perspective, we presented the principles, recent successes, and possible future applications of a powerful new approach to study disordered many-body systems through the estimation of the number and volume of basins of attractions.
This approach enables the computation of the Shannon entropy, that appears as a natural extension of the thermodynamic entropy of equilibrium systems, in a much broader class of systems.
It has already been shown that basin-volume computations work in the specific case of granular packings, but there is an avenue ahead.
Not only can this method lead to estimations of the features of complex energy landscapes in thermal systems like supercooled liquids, molecular liquids or spin glasses, it also provides a systematic and quantitative way of characterising the state space of generic dynamical systems.
At a time when numerical and experimental studies of a broad spectrum of non-equilibrium many-body systems are booming, in spite of the absence of a general theoretical framework to understand them, such a generic tool could prove to be invaluable.
Our vision is that the theoretical ideas and protocols presented herein will become new canonical forms of computation adopted in multiple areas of science and engineering.

\begin{acknowledgements}
We wish to thank Daniel Asenjo, Daan Frenkel, Johannes Gasteiger (né Klicpera), Fabien Paillusson, K. Julian Schrenk, Jacob Stevenson, and Praharsh Suryadevara for many invaluable contributions to the theoretical underpinnings and computer implementation of the methods described in this perspective. The authors acknowledge funding from the Simons Foundation through the Simons Center for Computational Physical Chemistry, Department of Chemistry, NYU. S. M. gracefully acknowledges support from the National Science Foundation grant IIS-2226387. S. M. performed part of this work at the Aspen Center for Physics, which is supported by National Science Foundation grant PHY-1607611.
\end{acknowledgements}

\appendix
\section{Scaling of the number of replicas \label{app:replicaspacing}}

We derive a scaling for the required number of replicas, $K_d$, to correctly estimate the volume of a hypercube in dimension $d$.
Each of these replicas is attached to the center of the cube by a spring with rigidity $k$, and samples positions according to a distribution that is well approximated by a $d-$dimensional Gaussian.
The narrowest of these distributions is the one for $k_{max}$, $h_{k_{max}}$(r), with variance
\begin{align}
    \sigma_{max}^2 = \frac{1}{k_{max}}\left( d - 2\frac{\Gamma\left(\frac{d+1}{2}\right)}{\Gamma\left(\frac{d}{2}\right)}\right).
\end{align}
We can impose that the mode of $h_{k_{max}}$ coincides with the radius, $a$, of the largest hyperball inscribed in the hypercube, yielding the condition
\begin{align}
k_{max} = \frac{d-1}{a^2}.
\label{eq:kmax}
\end{align}
Finally, we know that the $k=0$ replica will have most of its mass concentrated at $a\sqrt{d/3}$. We can therefore require that the number replicas, $K_d$, is such that the interval $[a; \sqrt{d/3}a]$ sums up to $K_d \times 2\sigma_{max}$.
Putting all of this together, we arrive to the result:
\begin{align}
    K_d = \frac{\sqrt{3d} - 3}{6 \sqrt{\frac{d - 2\frac{\Gamma\left(\frac{d+1}{2}\right)}{\Gamma\left(\frac{d}{2}\right)}}{d - 1}}},
\end{align}
which for $d \to \infty$ reduces to $K_d \sim d / \sqrt{6}$.
In other words, in the case of a hypercube, the number of replicas necessary to estimate the volume grows linearly with $d$.

\section{Implications of the power-law between basin volumes and pressures \label{app:powerlaw}}

We report how to deduce an expression for the difference between the Boltzmann and Gibbs entropies of granular packings from the power-law scaling between the volumes of basins and the pressures of the corresponding packings.
Recall that the Gibbs entropy of a jammed configuration of $N$ particles can be written as $S_G = - \sum_i^\Omega p_i \ln p_i - \ln N!$ with a sum running over basins.
Noticing that $p_i = v_i / \mathcal{V}$, one gets $S_G = - \langle \ln v_i \rangle_{B} - \ln N! + \ln\mathcal{V}$, where the average is biased proportionally to the volume of the basins.
The power-law relation shown in Fig.~\ref{fig:granularresults}$b$ entails a linear law between logs, $- \langle \ln v_i \rangle_{B} = N(\lambda(\phi) \langle \ln P  \rangle_{B} + c(\phi))$.
Therefore, the Gibbs entropy reads 
\begin{equation}
S_G = N\lambda \langle \ln P  \rangle_{B} + Nc - \ln N! + \ln \mathcal{V}.
\label{eq:granular_gibbs}
\end{equation}
Now, the Boltzmann entropy at a given value of pressure and volume can be computed as the log of the number of configurations $\Omega(P,V)$, which can be written as
\begin{align}
    \Omega(P,V) &= \Omega(V) \int\limits_{P}^{P + \delta P} U(\Pi|V) \mathrm{d}\Pi,
\end{align}
where $U(P|V)$ is the unbiased distribution of pressures given a total volume $V$, and the integral over a small element imposes the value of $P$ within the measure $U$.
Since in our protocol the basins are sampled proportionally to their volume, $B$ is linked to $U$ via
\begin{align}
    B(P|V) = U(P|V) \frac{v(P, \phi, N)}{\langle v \rangle},
\end{align}
with $v(P, \phi, N)$ the volume of a basin at pressure $P$, packing fraction $\phi$ and $N$ particles.
According to the results in Fig.~\ref{fig:granularresults}$b$, these volumes verify $v(P,\phi,N) = P^{-N \lambda} e^{-N c}$.
Recalling that $\Omega(V) = \mathcal{V}/\langle v\rangle$, the number of configurations with a fixed pressure reads
\begin{align}
    \Omega(P,V) &= \mathcal{V} e^{N c}\int\limits_{P}^{P + \delta P} B(\Pi|V) \Pi^{N\lambda} \mathrm{d}\Pi.
\end{align}
Furthermore, the number of configurations considering all possible pressures at a given volume is simply the sum
\begin{align}
    \Omega(V) = \int\limits_{0}^\infty \mathrm{d}P \Omega(P,V).
\end{align}
The Boltzmann entropy at a given volume therefore reads
\begin{align}
    S_B(V) = &\ln \left[ \int\limits_{0}^\infty \mathrm{d}P  \int\limits_{P}^{P + \delta P} B(\Pi|V) \Pi^{N \lambda} d\Pi\right] \nonumber \\
    &+ N c - \ln N! + \ln \mathcal{V}.
    \label{eq:granular_boltz}
\end{align}
In order to simplify this expression, we use the result of Ref.~\cite{Martiniani2017a} that $\ln P$ is approximately normal-distributed, so that
\begin{align}
    B(P|V) P^{N \lambda} \approx \frac{1}{P \sqrt{2 \pi \sigma^2}} e^{ - \frac{(\ln P - \mu)^2}{2 \sigma^2} + N \lambda \ln P}.
\end{align}
It is convenient to rewrite the integral in square brackets in Eq.~\ref{eq:granular_boltz} as
\begin{align}
    \Xi &= \int\limits_{0}^{\infty} dP \int\limits_{0}^{\infty} d\Pi B(\Pi|V) x^{N \lambda} \delta(\Pi - P).
\end{align}
After some algebra, one finds
\begin{align}
    \Xi = \exp\left[N \lambda \mu + N^2 \lambda^2 \frac{\sigma^2}{2}\right].
\end{align}
Using the empirical observation that $\sigma^2 = s / N$ and $\mu = \mu_\infty + m / N$ where $s, \mu_\infty, m$ are all $O(1)$~\cite{Martiniani2017a}, we find that
\begin{align}
    \lim_{N\to\infty} \frac{\ln \Xi}{N}= \lambda \mu_\infty + \lambda^2 \frac{s}{2}.
    \label{eq:xi_lim}
\end{align}
Identifying $\mu_{\infty}$ with $\langle \ln P \rangle_{B}$, from Eqs.~\ref{eq:granular_gibbs}, \ref{eq:granular_boltz} and \ref{eq:xi_lim} we finally get
\begin{align}
   \lim_{N\to\infty}\frac{1}{N}\left[S_B(V) - S_G(V)\right] = \lambda^2 \frac{s}{2}.
\end{align}
This expression shows that the two entropies only coincide when $\lambda \to 0$, but also that $S_B \geq S_G$, as observed in the measurements reported in Fig.~ \ref{fig:granularresults} of the main text.


\begin{thebibliography}{0}%
\makeatletter
\providecommand \@ifxundefined [1]{%
 \@ifx{#1\undefined}
}%
\providecommand \@ifnum [1]{%
 \ifnum #1\expandafter \@firstoftwo
 \else \expandafter \@secondoftwo
 \fi
}%
\providecommand \@ifx [1]{%
 \ifx #1\expandafter \@firstoftwo
 \else \expandafter \@secondoftwo
 \fi
}%
\providecommand \natexlab [1]{#1}%
\providecommand \enquote  [1]{``#1''}%
\providecommand \bibnamefont  [1]{#1}%
\providecommand \bibfnamefont [1]{#1}%
\providecommand \citenamefont [1]{#1}%
\providecommand \href@noop [0]{\@secondoftwo}%
\providecommand \href [0]{\begingroup \@sanitize@url \@href}%
\providecommand \@href[1]{\@@startlink{#1}\@@href}%
\providecommand \@@href[1]{\endgroup#1\@@endlink}%
\providecommand \@sanitize@url [0]{\catcode `\\12\catcode `\$12\catcode
  `\&12\catcode `\#12\catcode `\^12\catcode `\_12\catcode `\%12\relax}%
\providecommand \@@startlink[1]{}%
\providecommand \@@endlink[0]{}%
\providecommand \url  [0]{\begingroup\@sanitize@url \@url }%
\providecommand \@url [1]{\endgroup\@href {#1}{\urlprefix }}%
\providecommand \urlprefix  [0]{URL }%
\providecommand \Eprint [0]{\href }%
\providecommand \doibase [0]{http://dx.doi.org/}%
\providecommand \selectlanguage [0]{\@gobble}%
\providecommand \bibinfo  [0]{\@secondoftwo}%
\providecommand \bibfield  [0]{\@secondoftwo}%
\providecommand \translation [1]{[#1]}%
\providecommand \BibitemOpen [0]{}%
\providecommand \bibitemStop [0]{}%
\providecommand \bibitemNoStop [0]{.\EOS\space}%
\providecommand \EOS [0]{\spacefactor3000\relax}%
\providecommand \BibitemShut  [1]{\csname bibitem#1\endcsname}%
\let\auto@bib@innerbib\@empty
\end{thebibliography}%


\begin{thebibliography}{115}

\bibitem{Attard2002} P Attard, \textit{Thermodynamics and statistical mechanics: equilibrium by entropy maximisation}, Academic Press, London (2002).
\bibitem{Gibbs1902} J W Gibbs, \textit{Elementary principles in statistical mechanics}, Charles Scribner's Sons, New York (1902).
\bibitem{Kardar2007} M Kardar, \textit{Statistical Physics of Particles}, Cambridge University Press (2007).
\bibitem{Carnot1824} S Carnot, \textit{Réflexions sur la puissance motrice du feu et sur les machines propres à développer cette puissance}, Bachelier (1824).
\bibitem{Martiniani2017} S Martiniani, \textit{On the complexity of energy landscapes: algorithms and a direct test of the Edwards conjecture}, University of Cambridge (2017).
\bibitem{Cover1999} T M Cover, \textit{Elements of information theory}, John Wiley \& Sons (1999).
\bibitem{Frenkel2001} D Frenkel, B Smit, \textit{Understanding molecular simulation: from algorithms to applications, Vol. I}, Elsevier (2001).
\bibitem{LandauBinder} D Landau, K Binder, \textit{A Guide to Monte Carlo Simulations in Statistical Physics, Second Edition}, Cambridge University Press (2005). 
\bibitem{Wales2003} D Wales, \textit{Energy landscapes: Applications to clusters, biomolecules and glasses}, Cambridge University Press (2003).
\bibitem{Stillinger2015} F H Stillinger, \textit{Energy landscapes, inherent structures, and condensed-matter phenomena}, Princeton University Press (2015).
\bibitem{Kostlan2002} E Kostlan, \textit{On the expected number of real roots of a system of random polynomial equations}, Foundations of Computational Mathematics, World Scientific (2002).
\bibitem{Lotka1925} A J Lotka, \textit{Elements of Physical Biology}, Williams and Wilkins (1925).
\bibitem{Artstein-Avidan2015} S Artstein-Avidan, A Giannopoulos, V D Milman, \textit{Asymptotic Geometric Analysis, Part I}, Mathematical Surveys and Monographs, Volume 202, American Mathematical Society, Providence, Rhode Island (2015).
\bibitem{Gelfand} I Gel'Fand, G Shilov, \textit{Generalized functions, Vol. 1}, Academic Press (1968).

\bibitem{Clausius1854} R Clausius, \textit{Über eine veränderte Form des zweiten Hauptsatzes der mechanischen Wärmetheorie}, Annalen der Physik. \textbf{xciii} (12) 481–506 (1854).
\bibitem{Shannon1948} C E Shannon, \textit{A mathematical theory of communication}, The Bell system technical journal, \textbf{27}(3):379--423 (1948).
\bibitem{Martiniani2019} S Martiniani, P M Chaikin, D Levine, \textit{Quantifying hidden order out of equilibrium}, Phys. Rev. X \textbf{9}, 011031 (2019).
\bibitem{Martiniani2020} S Martiniani, Y Lemberg, P M Chaikin, D Levine, \textit{Correlation lengths in the language of computable information}, Phys. Rev. Lett. \textbf{125}, 170601 (2020).
\bibitem{Avinery2019} R Avinery, M Kornreich, R Beck, \textit{Universal and Accessible Entropy Estimation Using a Compression Algorithm}, Phys. Rev. Lett. \textbf{123}, 178102 (2019).
\bibitem{cavagna2021vicsek} A Cavagna, P M Chaikin, D Levine, S Martiniani, A Puglisi, M Viale, \textit{Vicsek model by time-interlaced compression: A dynamical computable information density}, Phys. Rev. E \textbf{103}, 062141 (2021).
\bibitem{ro2022model} S Ro, B Guo, A Shih, T V Phan, R H Austin, D Levine, P M Chaikin, and S Martiniani, \textit{Model-Free Measurement of Local Entropy Production and Extractable Work in Active Matter}, Phys. Rev. Lett. \textbf{129}, 220601 (2022).
\bibitem{Wiley2006} D A Wiley, S H Strogatz, M Girvan, \textit{The size of a sync basin}, Chaos, \textbf{16}(1), 015103 (2006).
\bibitem{Cornelius2013} S P Cornelius, W L Kath, A E Motter, \textit{Realistic control of network dynamics}, Nat. Commun. \textbf{4}(1):1--9 (2013).
\bibitem{Zhang2021} Y Zhang, S H Strogatz, \textit{Basins with tentacles}, Phys. Rev. Lett., \textbf{127}, 194101 (2021).
\bibitem{Frenkel1984} D Frenkel, A J Ladd, \textit{New Monte Carlo mehod to compute the free energy of arbitrary solids. Application to the fcc and hcp phases of hard spheres}, J. Chem. Phys., \textbf{81}(7), 3188 (1984).
\bibitem{Xu2011} N Xu, D Frenkel, A J Liu, \textit{Direct determination of the size of basins of attraction of jammed solids}, Phys. Rev. Lett., \textbf{106}(24), 245502 (2011).
\bibitem{Asenjo2014} D Asenjo, F Paillusson, D Frenkel, \textit{Numerical calculation of granular entropy}, Phys. Rev. Lett., \textbf{112}(9), 098002 (20114).
\bibitem{Martiniani2017a} S Martiniani, K J Schrenk, K Pamola, B Chakraborty, D Frenkel, \textit{Numerical test of the Edwards conjecture shows that all packings are equally probable at jamming}, Nat. Phys., \textbf{13}, 848 (2017).
\bibitem{Martiniani2016} S Martiniani, K J Schrenk, J D Stevenson, D J Wales, D Frenkel, \textit{Turning intractable counting into sampling: computing the configurational entropy of three-dimensional jammed packings}, Phys. Rev. E, \textbf{93}(1), 012906 (2016).
\bibitem{Martiniani2016a} S Martiniani, K J Schrenk, J D Stevenson, D J Wales, D Frenkel, \textit{Structural analysis of high-dimensional basins of attraction}, Phys. Rev. E, \textbf{93}(3), 031301 (2016).
\bibitem{Frenkel2017} D Frenkel, K J Schrenk, S Martiniani, \textit{Monte Carlo sampling for stochastic weight functions}, Proc. Nat. Ac. Sci., \textbf{114}(27), 6924 (2017).
\bibitem{Ashwin2012} S S Ashwin, J Blawzdziewicz, C S O'Hern, M D Shattuck, \textit{Calculations of the structure of basin volumes for mechanically stable packings}, Phys. Rev. E, \textbf{85}, 061307 (2012).
\bibitem{Shirts2008} M R Shirts, J D Chodera, \textit{Statistically optimal analysis of samples from multiple equilibrium states}, J. Chem. Phys., \textbf{129}(12), 124105 (2008).
\bibitem{Ding2019} X Ding, J Vilseck, C Brooks, \textit{Fast Solver for Large Scale Multistate Bennett Acceptance Ratio Equations}, J. Chem. Theory Comput., \textbf{2019}(15), 799--802 (2019).
\bibitem{Kirkwood1935} J G Kirkwood, \textit{Statistical Mechanics of Fluid Mixtures}, J. Chem. Phys. \textbf{3}, 300 (1935).
\bibitem{Gelman1998} A Gelman, X-L Meng, \textit{Simulating normalizing constants: from importance sampling to bridge sampling to path sampling}, Statist. Sci. \textbf{13} (2) 163--185 (1998).
\bibitem{Bunker2000} A Bunker, B D\"{u}nweg, \textit{Parallel excluded volume tempering for polymer melts}, Phys. Rev. E, \textbf{63}(1), 016701 (2000).
\bibitem{Fukunishi2002} H Fukunishi, O Watanabe, S Takada, \textit{On the hamiltonian replica exchange method for efficient sampling of biomolecular systems: Application to protein structure prediction}, J. Chem. Phys., \textbf{116}(20), 9058 (2002).
\bibitem{Torrie1977} J M Torrie, J P Valleau, \textit{Nonphysical sampling distributions in Monte Carlo free-energy estimation: Umbrella sampling}, J. Comput. Phys. \textbf{23} (2), 187--199 (1977).
\bibitem{Chevallier2022} A Chevallier, F Cazals, P Fearnhead, \textit{Efficient computation of the volume of a polytope in high dimensions using Piecewise Deterministic Markov Processes}, ArXiv Preprint, 2202.09129 (2022).
\bibitem{Skilling2012} J Skilling, \textit{Bayesian Computation in big spaces-nested sampling and Galilean Monte Carlo}, AIP Conference Proceedings \textbf{1443}, 145 (2012).
\bibitem{Griffiths2019} M Griffiths, D J Wales, \textit{Nested Basin-Sampling}, J. Chem. Theory Comput., \textbf{15}(12),  6865--6881 (2019).
\bibitem{Edwards1989} S F Edwards, R B S Oakeshott, \textit{Theory of Powders}, Physica A, \textbf{157}(3), 1080 (1989).
\bibitem{Baule2018} A Baule, F Morone, H J Herrmann, H A Makse, \textit{Edwards statistical mechanics for jammed granular matter}, Rev. Mod. Phys., \textbf{90}(1), 015006 (2018).
\bibitem{Aste2006} T Aste, \textit{Volume fluctuations and geometrical constraints in granular packs}, Phys. Rev. Lett., \textbf{96}(1), 018002 (2006).
\bibitem{Aste2008} T Aste, T Di Matteo \textit{Emergence of Gamma distributions in granular materials and packing models}, Phys. Rev. E, \textbf{77}(2), 021309 (2008).
\bibitem{McNamara2009} S McNamara, P Richard, S K De Richter, G Le Ca\"{e}r, R Delannay \textit{Measurement of granular entropy}, Phys. Rev. E, \textbf{80}(3), 031301 (2009).
\bibitem{Puckett2013} J G Puckett, K E Daniels \textit{Equilibrating temperature-like variables in jammed granular subsystems}, Phys. Rev. Lett., \textbf{110}(5), 058001 (2013).
\bibitem{Zhao2014} S-C Zhao, M Schr\"{o}ter, \textit{Measuring the configurational entropy of a binary disc packing}, Soft Matter, \textbf{10}(23), 4208 (2014).
\bibitem{Bililign2019} E S Bililign, J E Kollmer, K E Daniels \textit{Protocol dependence and state variables in the force-moment ensemble}, Phys. Rev. Lett., \textbf{122}(3), 038001 (2019).
\bibitem{Sun2020} X Sun, W Kob, R Blumenfeld, H Tong, Y Wang, J Zhang, \textit{Friction-controlled entropy-stability competition in granular systems}, Phys. Rev. Lett., \textbf{125}(26), 268005 (2020).
\bibitem{Yuan2021} Y Yuan, Y Xing, J Zheng, Z Li, H Yuan, S Zhang, Z Zeng, C Xia, H Tong, W Kob, J Zhang, Y Wang, \textit{Experimental test of the Edwards volume ensemble for tapped granular packings}, Phys. Rev. Lett., \textbf{127}(1), 018002 (2021).
\bibitem{Weeks1971} J D Weeks, D Chandler, H C Andersen, \textit{Role of repulsive forces in determining the equilibrium structure of simple liquids}, J. Chem. Phys. \textbf{54}, 5237 (1971).
\bibitem{Goodrich2014} C P Goodrich, S Dagois-Bohy, B P Tighe, M Van Hecke, A J Liu, S R Nagel, \textit{Jamming in finite systems: Stability, anisotropy, fluctuations, and scaling}, Phys. Rev. Lett., \textbf{112}(14), 145502 (2014).
\bibitem{Atkinson2013} S Atkinson, F Stillinger, S Torquato, \textit{Detailed characterization of rattlers in exactly isostatic, strictly jammed sphere packings}, Phys. Rev. E, \textbf{88}(6), 062208 (2013).
\bibitem{Gao2006} G-J Gao, J B\l{}awzdziewicz, C S O'Hern, \textit{Frequency distribution of mechanically stable disk packings}, Phys. Rev. E \textbf{74}, 061304 (2006).
\bibitem{Parrinello1980} M Parrinello, A Rahman, \textit{Crystal Structure and Pair Potentials: A Molecular-Dynamics Study}, Phys. Rev. Lett. \textbf{45}, 413--420 (1980).
\bibitem{Parrinello1981} M Parrinello, A Rahman, \textit{Polymorphic transitions in single crystals: A new molecular dynamics method}, J. App. Phys. \textbf{52}, 7182--7190 (1981).
\bibitem{Parrinello1982} M Parrinello, A Rahman, \textit{Strain fluctuations and elastic constants}, J. Chem. Phys. \textbf{76}, 2662--2666 (1982).
\bibitem{Najafabadi1983} R Najafabadi, S Yip, \textit{Observation of Finite-Temperature Bain Transformation (f.c.c. $\leftrightarrow$ b.c.c.) in Monte Carlo Simulation of Iron}, Scr. Metall. \textbf{17}, 1199--1204 (1983).
\bibitem{Yashonath1985} S Yashonath, C Rao, \textit{A monte carlo study of crystal structure transformations}, Mol. Phys. \textbf{54}, 245--251 (1985).
\bibitem{Baldock2016} R J N Baldock, L B P\'{a}rtay, A P Bart\'{o}k, M C Payne, G Cs\'{a}nyi, \textit{Determining pressure-temperature phase diagrams of materials}, Phys. Rev. B \textbf{93}, 174108 (2016).
\bibitem{Baldock2017} R J N Baldock, N Bernstein, K Michael Salerno, L B P\'{a}rtay, G Cs\'{a}nyi, \textit{Constant-pressure nested sampling with atomistic dynamics}, Phys. Rev. E \textbf{96}, 043311 (2017).
\bibitem{Anzivino2022} C Anzivino, M Casiulis, T Zhang, A S Moussa, S Martiniani, A Zaccone, \textit{Estimating random close packing in polydisperse and bidisperse hard spheres via an equilibrium model of crowding}, J. Chem. Phys., In Press (2022). 
\bibitem{Castellani2005} T Castellani, A Cavagna, \textit{Spin-glass theory for pedestrians}, J. Stat. Mech., \textbf{2005}, P05012 (2005).
\bibitem{Ros2019} V Ros, G Ben Arous, G Biroli, C Cammarota, \textit{Complex Energy Landscapes in Spiked-Tensor and Simple Glassy Models:
Ruggedness, Arrangements of Local Minima, and Phase Transitions}, Phys. Rev. X, \textbf{9}(1), 011003 (2019).
\bibitem{Rizzo2021} T Rizzo, \textit{Path integral approach unveils role of complex energy landscape for activated dynamics of glassy systems}, Phys. Rev. B \textbf{104}, 094203 (2021).
\bibitem{Folena2022} G Folena, A Manacorda, F Zamponi, \textit{Introduction to the dynamics of disordered systems: equilibrium and gradient descent}, Lectures Notes for the Fundamental Problems in Statistical Physics XV Summer School (2021).
\bibitem{Ros2019b} V Ros, G Biroli, C Cammarota, \textit{Complexity of energy barriers in mean-field glassy systems}, EPL \textbf{126}, 20003 (2019).
\bibitem{Ros2021} V Ros, G Biroli, C Cammarota, \textit{Dynamical Instantons and Activated Processes in Mean-Field Glass Models}, SciPost Phys. \textbf{10}, 002 (2021).
\bibitem{LacroixAChezToine2022} B Lacroix-à-chez-Toine, Y Fyodorov, \textit{Counting equilibria in a random non-gradient dynamics with heterogeneous relaxation rates},  J. Phys. A: Math. Theor. \textbf{55}, 144001 (2022).
\bibitem{KentDobias2021} J Kent-Dobias, J Kurchan, \textit{Complex complex landscapes}, Phys. Rev. Research \textbf{3}, 023064 (2021).
\bibitem{Bitzek2006} E Bitzek, P Koskinen, F G\"{a}hler, M Moseler, P Gumbsch, \textit{Structural relaxation made simple}, Phys. Rev. Lett., \textbf{97}(17), 170201 (2006).
\bibitem{Stillinger1982} F H Stillinger, T A Weber, \textit{Hidden structure in liquids}, Phys. Rev. A, \textbf{25}(2), 978 (1982).
\bibitem{Cohen1996} S D Cohen, A C Hindmarsh, P F Dubois, \textit{ CVODE, a stiff/nonstiff ODE solver in C}, Computers in Physics, \textbf{10}(2), 138 (1996).
\bibitem{Charbonneau2017} P Charbonneau, J Kurchan, G Parisi, P Urbani, F Zamponi, \textit{Glass and jamming transitions: From exact results to finite-dimensional descriptions}, Ann. Rev. Cond. Mat. Phys., \textbf{8}(1), 265 (2017).
\bibitem{Franz1995} S Franz, G Parisi,  \textit{Recipes for metastable states in spin glasses}, Journal de Physique I, \textbf{5}(11), 1401 (1995).
\bibitem{Parisi2010} G Parisi, F Zamponi, \textit{Mean-field theory of hard sphere glasses and jamming}, Rev. Mod. Phys., \textbf{82}(1), 789 (2010).
\bibitem{Berthier2019} L Berthier, M Ozawa, C Scalliet, \textit{Configurational entropy of glass-forming liquids}, J. Chem. Phys., \textbf{150}(16), 160902 (2019).
\bibitem{Rulquin2016} C Rulquin, P Urbani, G Biroli, G Tarjus, M Tarzia, \textit{Nonperturbative fluctuations and metastability in a simple model: from observables to microscopic theory and back}, J. Stat. Mech., \textbf{2016}(2), 023209 (2016).
\bibitem{Stillinger1995} F H Stillinger, \textit{A topographic view of supercooled liquids and glass formation}, Science, \textbf{267}(5206), 1935 (1995).
\bibitem{Sciortino2005} F Sciortino, \textit{Potential energy landscape description of supercooled liquids and glasses}, J. Stat. Mech., \textbf{2005}(5), P05015 (2005).
\bibitem{Berthier2017} L Berthier, P Charbonneau, D Coslovich, A Ninarello, M Ozawa, S Yaida, \textit{Configurational entropy measurements in extremely supercooled liquids that break the glass ceiling}, Proc. Nat. Ac. Sci., \textbf{114}(43), 11356 (2017).
\bibitem{Ozawa2018} M Ozawa, G Parisi, L Berthier, \textit{Configurational entropy of polydisperse supercooled liquids}, J. Chem. Phys., \textbf{149}(15), 154501 (2018).
\bibitem{Nishikawa2022} Y Nishikawa, M Ozawa, A Ikeda, P Chaudhuri, L Berthier, \textit{Relaxation Dynamics in the Energy Landscape of Glass-Forming Liquids}, Phys. Rev. X, \textbf{12}, 021001 (2022). 
\bibitem{Scalliet2021} C Scalliet, B Guiselin, L Berthier, \textit{Excess wings and asymmetric relaxation spectra in a facilitated trap model}, J. Chem. Phys., \textbf{155}, 064505 (2021). 
\bibitem{Guiselin2022} B Guiselin, C Scalliet, L Berthier, \textit{Microscopic origin of excess wings in relaxation spectra of supercooled liquids}, Nat. Phys., \textbf{18}, 468--472 (2022). 
\bibitem{Chakraborty2011} A Chakraborty, P Seiler, G J Balas, \textit{Susceptibility of F/A-18 flight controllers to the fallingleaf mode: Nonlinear analysis}, J. Guid. Control Dyn., \textbf{34}(1), 73 (2011).
\bibitem{Kac1943} M Kac, \textit{On the average number of real roots of a random algebraic equation}, Bull. Am. Math. Soc., \textbf{49}(4), 314 (1943).
\bibitem{Farahmand1986} K Farahmand, \textit{On the average number of real roots of a random algebraic equation}, Ann. Probab., \textbf{14}(2), 702 (1986).
\bibitem{Bogomolny1992} E Bogomolny, O Bohigas, P Leboeuf, \textit{Distribution of roots of random polynomials}, Phys. Rev. Lett., \textbf{68}(18), 2726 (1992).
\bibitem{Edelman1995} A Edelman, E Kostlan, \textit{How many zeros of a random polynomial are real?}, Bull. Am. Math. Soc., \textbf{32}(1), 1 (1995).
\bibitem{Rojas1996} J M Rojas, \textit{On the average number of real roots of certain random sparse polynomial systems}, Lectures in Applied Mathematics-American Mathematical Society, \textbf{32}, 689 (1996).
\bibitem{Fyodorov2004} Y V Fyodorov, \textit{Complexity of random energy landscapes, glass transition, and absolute value of the spectral determinant of random matrices}, Phys. Rev. Lett., \textbf{92}(24), 240601 (2004).
\bibitem{Malajovich2004} G Malajovich, J M Rojas, \textit{High probability analysis of the condition number of sparse polynomial systems}, Theor. Comput. Sci., \textbf{315}(2-3), 525 (2004).
\bibitem{Azais2005} J-M Aza\"{i}s, M Wschebor, \textit{On the roots of a random system of equations. The theorem of Shub and Smale and some extensions}, Found. Comput. Math., \textbf{5}(2), 125 (2005).
\bibitem{Armentano2009} D Armentano, M Wschebor, \textit{Random systems of polynomial equations. The expected number of roots under smooth analysis}, Bernoulli, \textbf{15}(1), 249 (2009).
\bibitem{Fyodorov2012} Y V Fyodorov, G A Hiary, J Keating, \textit{Freezing transition, characteristic polynomials of random matrices, and the Riemann zeta function}, Phys. Rev. Lett., \textbf{108}(17), 170601 (2012).
\bibitem{Fyodorov2012a} Y V Fyodorov, C Nadal, \textit{Critical behavior of the number of minima of a random landscape at the glass transition point and the Tracy-Widom distribution}, Phys. Rev. Lett., \textbf{109}(16), 167203 (2012).
\bibitem{Fyodorov2013} Y V Fyodorov, \textit{High-dimensional random fields and random matrix theory}, ArXiv Preprint, arXiv:1307.2379 (2013).
\bibitem{Fyodorov2015} Y V Fyodorov, A Lerario, E Lundberg, \textit{On the number of connected components of random algebraic hypersurfaces}, J. Geom. Phys., \textbf{95}, 1 (2015).
\bibitem{Cheng2015} D Cheng, A Schwartzman, \textit{On the explicit height distribution and expected number of local maxima of isotropic Gaussian random fields}, ArXiv Preprint, arXiv:1503.01328 (2015).
\bibitem{Krzakala2007} F Krzakala, J Kurchan, \textit{Landscape analysis of constraint satisfaction problems}, Phys. Rev. E \textbf{76}, 021122 (2007).
\bibitem{Krzakala2008} F Krzakala, L Zdeborová, \textit{Hiding Quiet Solutions in Random Constraint Satisfaction Problems}, Phys. Rev. Lett. \textbf{102}, 238701 (2009).
\bibitem{Zdeborova2008} L Zdeborová, M Mézard, \textit{Constraint Satisfaction Problems with Isolated Solutions are Hard}, J. Stat. Mech. \textbf{2008}, P12004 (2008).
\bibitem{TayaraniNarajan2013} M-H Tayarani-Narajan, A Prügel-Bennett, \textit{On the Landscape of Combinatorial Optimisation Problems}, IEEE Transactions on Evolutionary Computation, \textbf{18} (3), 420-434 (2013).
\bibitem{Bachlechner2019} T C Bachlechner, K Eckerle, O Janssen, M Kleban, \textit{Axion landscape cosmology}, JCAP09(\textbf{2019})062
\bibitem{Lotka1910} A J Lotka, \textit{Contribution to the Theory of Periodic Reaction}, J. Phys. Chem. \textbf{14}(3), 271 (1910).
\bibitem{Lotka1920} A J Lotka, \textit{Analytical Note on Certain Rhythmic Relations in Organic Systems}, Proc. Natl. Acad. Sci. U.S.A. \textbf{6}(7), 410 (1920).
\bibitem{Volterra1926} V Volterra, \textit{Variazioni e fluttuazioni del numero d'individui in specie animali conviventi}, Mem. Acad. Lincei Roma \textbf{2}, 31 (1926).
\bibitem{Bunin2017} G Bunin, \textit{Ecological communities with Lotka-Volterra dynamics}, Phys. Rev. E \textbf{95}(4), 042414 (2017).
\bibitem{Biroli2018} G Biroli, G Bunin, C Cammarota, \textit{Marginally stable equilibria in critical ecosystems}, New J. Phys. \textbf{20}, 083051 (2018).
\bibitem{Roy2019} F Roy, G Biroli, G Bunin, C Cammarota, \textit{Numerical implementation of dynamical mean field theory for disordered systems: application to the Lotka–Volterra model of ecosystems}, J. Phys. A: Math. Theor. \textbf{52}, 484001 (2019).
\bibitem{Altieri2021} A Altieri, F Roy, C Cammarota, G Biroli, \textit{Properties of Equilibria and Glassy Phases of the Random Lotka-Volterra Model with Demographic Noise}, Phys. Rev. Lett. \textbf{126}, 258301 (2021).
\bibitem{Barbier2018} M Barbier, J-F Arnoldi, G Bunin, M Loreau, \textit{Generic assembly patterns in complex ecological communities}, Proc. Natl. Acad. Sci. U.S.A. \textbf{115}(9), 2156 (2018).
\bibitem{Bunin2018} G Bunin, \textit{Directionality and community-level selection}, Oikos \textbf{130}(4), 489 (2018).
\bibitem{Barbier2021} M Barbier, C de Mazancourt, M Loreau, G Bunin, \textit{Fingerprints of High-Dimensional Coexistence in Complex Ecosystems}, Phys. Rev. X \textbf{11}(1), 011009 (2021).
\bibitem{ballard2017energy} A J Ballard, R Das, S Martiniani, D Mehta, L Sagun, J D Stevenson, D J Wales \textit{Energy landscapes for machine learning}, Phys. Chem. Chem. Phys., \textbf{19}, 12585-12603 (2017).
\bibitem{Hochreiter1994} S Hochreiter, J Schmidhuber, \textit{Simplifying neural nets by discovering flat minima}, NeurIPS \textbf{7} (1994).
\bibitem{Hochreiter1997} S Hochreiter, J Schmidhuber, \textit{Flat minima}, Neur. Comput. \textbf{9}(1), 1 (1997).
\bibitem{Baldassi2020} C Baldassi, F Pittorino, R Zecchina, \textit{Shaping the learning landscape in neural networks around wide flat minima}, Proc. Natl. Acad. Sci. U.S.A. \textbf{117}(1), 161 (2020).
\bibitem{Feng2021} Y Feng, Y Tu, \textit{The inverse variance–flatness relation in stochastic gradient descent is critical for finding flat minima}, Proc. Natl. Acad. Sci. U.S.A. \textbf{118}(9), e2015617118 (2021).
\bibitem{Zhang2021a} S Zhang, I Reid, G P\'{e}rez, A Louis, \textit{Why Flatness Correlates With Generalization For Deep Neural Networks}, Arxiv Preprint arXiv:2103.06219 (2021).
\bibitem{Pittorino2021} F Pittorino, C Lucibello, C Feinauer, G perugini, C Baldassi, E Demyanenko, R Zecchina, \textit{Entropic gradient descent algorithms and wide flat minima}, J. Stat. Mech. \textbf{2021}, 124015 (2021).
\bibitem{Mannelli2020} S S Mannelli, G Biroli, C Cammarota, F Krzakala, P Urbani, L Zdeborov\'{a}, \textit{Complex dynamics in simple neural networks: Understanding gradient flow in phase retrieval}, NeurIPS \textbf{33}, 3265 (2020).
\bibitem{Ceperley1999} D M Ceperley, M Dewing, \textit{The penalty method for random walks with uncertain energies}, J. Chem. Phys. \textbf{110}(20), 9812 (1999).
\bibitem{Stratonovich1957} R L Stratonovich, \textit{On a Method of Calculating Quantum Distribution Functions}, Soviet Physics Doklady, \textbf{2}, 416 (1957).
\bibitem{Hubbard1959} J Hubbard, \textit{Calculation of Partition Functions}, Phys. Rev. Lett. \textbf{3}, 77 (1959).


\bibitem{MathworldGabrielsHorn}  E W Weisstein, \textit{Gabriel's Horn}, In: MathWorld--A Wolfram Web Resource, mathworld.wolfram.com/GabrielsHorn.html

\bibitem{frenkelpriv} D Frenkel, \textit{Private communication}.

\end{thebibliography}
\end{document}